\documentclass[]{pasj02} 
\usepackage{natbib} 
\usepackage{booktabs}   
\usepackage{amsmath}

\jyear{2026}
\Received{2026/02/10}
\Accepted{2026/04/09}

\begin{document} 

\title{The Impact of the Bar on Dense Gas and Star Formation in M83}

\author{
 Tianyi GU\textsuperscript{1}\orcid{0009-0006-7649-2509} \email{tianyigu2@gmail.com}
}
\author{
 Yoshimasa Watanabe \textsuperscript{1}\orcid{0000-0002-9668-3592} 
}
\altaffiltext{1}{Shibaura Institute of Technology, 3-7-5 Toyosu, Koto-ku, Tokyo 135-8548, Japan}



\maketitle

\begin{abstract}
Although the relationship between molecular gas content and the star formation rate (SFR) has been extensively studied in nearby galaxies, it remains controversial whether the star formation efficiency (SFE) depends on galactic structure. In particular, whether the SFE is suppressed in the bar region compared with other structures, and the physical origin of this suppression, remain poorly understood. In this study, we investigate variations in the SFE and its physical drivers in the bar region of the nearby spiral galaxy M83, using multi-wavelength observations each covering a single field of view ($\sim$60 arcsec) toward the bar and spiral arm regions on a scale of 200 pc observed with ALMA. From the molecular gas surface density derived from $^{12}$CO($J=2-1$) and $^{13}$CO($J=1-0$), the dense molecular gas surface density derived from HCN ($J = 1-0$), and the star formation rate surface density determined from extinction-corrected H$\alpha$, we find that the SFEs in the bar region are roughly a factor of two lower than those in the spiral arm, indicating that the SFE is systematically suppressed in the bar. Moreover, we find that the SFEs of dense gas are lower in the bar than in the arm by a factor of about 0.35. These results suggest that not only the efficiency of converting bulk molecular gas into stars reduced in the bar, but the efficiency of star formation from dense gas is also lower. In addition, the CO line widths are systematically larger in the bar region and exhibit a negative correlation with both the SFE and the dense-gas SFE, consistent with the interpretation that enhanced turbulent motions hinder star formation. Although the analysis is limited to small regions of M83, our results suggest that the suppression of the SFE is related to large-scale dynamical effects on the molecular gas, such as strong shocks induced by cloud–cloud collisions and/or shear, both driven by non-circular motions in the bar.
\end{abstract}


\section{Introduction}

Star formation in galaxies is a universal process, yet it exhibits remarkable diversity in its efficiency and modes. One of the primary factors responsible for this diversity are gas dynamics and environments associated with the galactic structures such as spiral arms, bars, and galactic centers. In particular, the presence of a bar with a non-axisymmetric gravitational potential is thought to strongly perturb gas flows and dynamical states of molecular gas, thereby significantly affecting star formation activities\citep[e.g.,][]{Tubbs1982, Fujimoto2014a, Fujimoto2020, Maeda2023}.

Previous observations of nearby galaxies have suggested that the star formation activities are often suppressed in bars. For examples, despite the presence of dust lanes along the bars, prominent H\,\textsc{ii} regions are not commonly observed in the bars of NGC~1300 and NGC~5383 \citep{Tubbs1982, Sheth2000}. Moreover, observational studies of molecular gas and star formation in barred galaxies (e.g., NGC 2903; \citep{Muraoka2016}, and NGC 4303; \citep{Yajima2019}) have shown that the star formation efficiency (SFE), defined as the star formation rate (SFR) per unit molecular gas mass, is systematically lower in the bars than in the spiral arms despite the presence of abundant molecular gas.

Several models for the SFE suppression mechanism have been proposed based on theoretical and numerical simulation studies. In the bars, the strong shock waves, large shear, and fast cloud-cloud collisions associated with non-circular motions driven by the bar potential are thought to excite molecular clouds into gravitationally unbound states and hinder star formation \citep[e.g.,][]{Athanassoula1992, Fujimoto2020}. For example, \citet{Fujimoto2014a, Fujimoto2020} demonstrated that cloud collision velocities increase in bar regions, consequently reducing SFE.

However, with the advancement of large-scale CO mapping surveys of nearby galaxies, this picture has become more complex. For example, statistical studies based on large CO mapping datasets observed with the Atacama Large Millimeter/submillimeter Array (ALMA) and the Institut de Radioastronomie Millimétrique (IRAM) 30m radio telescope suggested that the trend of SFE suppression in bar regions can disappear when averaging over galaxy samples, even though individual galaxies exhibit such suppression \citep{Querejeta2021, Diaz-Garcia2021}. \citet{Querejeta2021} analyzed PHANGS-ALMA data and concluded that there was no systematic differences in SFE among different structures, such as the center, bar, and arms, across 74 galaxies.  They also suggested that while galactic structures govern gas distribution, they may not strongly affect SFE itself.

The key to resolving these contradictory findings lies in the spatial scale of observations and region definition. Many statistical studies have treated multiple galaxies uniformly at relatively low resolution($15^{\prime\prime}$--$22^{\prime\prime}$)\citep{Querejeta2021, Diaz-Garcia2021, Muraoka2016}. At low angular resolution, data points classified as part of the bar may be contaminated by the regions with active star formation, such as the galactic center and the bar-end regions \citep{Maeda2023}. In some galaxies, the SFE values were overestimated in bar regions, due to contamination from the high SFE components in bar-end regions that exhibit enhanced star formation driven by gas flow convergence \citep{Renaud2015}.

Furthermore, recent studies focusing on dense molecular gas ($n_{\mathrm{H_2}} > 10^4$~cm$^{-3}$) have provided new insights into the relationship between star formation and environment. HCN ($J = 1-0$) is one of the dense molecular gas tracers observed in nearby galaxies as well as in high-redshift galaxies, due to its higher critical density. The luminosity of HCN ($J = 1-0$) is known to correlate strongly with the infrared luminosity, which serves as a tracer of SFR  \citep{Gao2004}. \citet{Neumann2024} observed HCN($J = 1-0$) in the barred spiral galaxy NGC 4321 with the ALMA on a scale of 260 pc and found that the SFR/HCN ratios, which is a proxy for the SFE of the dense gas(SFE$_{\mathrm{dense}}$), are systematically suppressed in the bar region. This result indicates that the correlation between dense gas and star formation is weaker in the bar than in the other regions. Moreover this implies that the suppression mechanism may work at at the stage of  star formation from the dense gas, rather than the accumulation of molecular gas. 

Our target galaxy M83 (NGC~5236) is an ideal target for detailed study of star formation activity, owing to its nearly face-on geometry with the inclination angle of 20$^{\circ}$ and its proximity (distance 4.5~Mpc; \citet{Thim2003}). Since the morphology of M83 is classified as SAB(s)c \citep{deVaucouleurs1991}, this galaxy possesses a prominent bar structure and spiral arms. Other fundamental parameters of M83 are summarized in Table~\ref{tab:m83_parameters}. The metallicity is comparable to or higher than that of the Milky Way \citep{Bresolin2016}. In such metal-rich environments, molecular clouds can be efficiently traced by CO emission lines. \citet{Handa1991} reported that the SFE is suppressed in the bar of M83 based on $^{12}$CO($J=1-0$) observation with the Nobeyama 45 m telescope and far-UV flux.

\begin{table}
  \tbl{Parameters of M83.}{%
  \begin{tabular}{lc}
      \hline
      Parameters & Value \\
      \hline
      Morphology\footnotemark[$*$] & SAB(s)c \\
      Center position\footnotemark[a] & R.A. (J2000): \timeform{13h37m00.72s} \\
          & Dec. (J2000): \timeform{-29D51'57.2''} \\
      Position angle\footnotemark[b] & \timeform{225D} \\
      Inclination angle\footnotemark[b] & \timeform{24D} \\
      Systemic velocity (LSR)\footnotemark[c] & $514$~$\mathrm{km~s^{-1}}$ \\
      Distance\footnotemark[d] & $4.5 \pm 0.3$~Mpc \\
      Linear scale\footnotemark[d] & $1^{\prime\prime} \sim 22$~pc \\
      SFR\footnotemark[e] & $3.0$~$\mathrm{M_{\odot}}$~yr$^{-1}$ \\
      H\,\textsc{i} mass\footnotemark[f] & $7.9 \times 10^9$~$\mathrm{M_{\odot}}$ \\
      H$_2$ mass\footnotemark[g] & $3.2 \times 10^9$~$\mathrm{M_{\odot}}$ \\
      \hline
    \end{tabular}}
    \label{tab:m83_parameters}
\begin{tabnote}
\footnotemark[$*$] \citet{deVaucouleurs1991}; 
\footnotemark[a] \citet{Thatte2000}; 
\footnotemark[b] \citet{Comte1981}; 
\footnotemark[c] \citet{Kuno2007}; 
\footnotemark[d] \citet{Thim2003}; 
\footnotemark[e] \citet{Jarrett2013}; 
\footnotemark[f] \citet{Heald2016}; 
\footnotemark[g] \citet{Schlegel1998}.
\end{tabnote}
\end{table}

This study aims to clarify the following two points through observations of bar region, part of the galactic center region, and spiral arm regions in M83.
\begin{enumerate}
    \item To verify whether the SFE and SFE$_{\mathrm{dense}}$ are indeed lower in the bar region than in the arm regions.
    \item If spatial differences in SFE and/or SFE$_{\mathrm{dense}}$ are confirmed, to investigate their physical origins.
\end{enumerate}

We used two molecular gas tracers, $^{12}$CO($J=2-1$) and $^{13}$CO($J=1-0$), for two reasons. First, while the optically thick $^{12}$CO($J=2-1$) line is excellent tracer of the large-scale distribution of molecular gas, it may not accurately estimate gas column density. In contrast, the $^{13}$CO($J=1-0$) line is relatively optically thin and is thought to more accurately reflect the column density of molecular gas. By comparing the two, we can more robustly assess the relationship between molecular gas mass and SFR. Second, the $^{12}$CO($J=2-1$) data observed with ALMA are becoming the standard for molecular gas studies in nearby galaxies (e.g., PHANGS-ALMA). Using this data allows for direct comparison and contrast of our results with other previous studies. Thus, by complementarily using the highly universal $^{12}$CO($J=2-1$) and the $^{13}$CO($J=1-0$), which is superior for estimating physical conditions, we aim to examine the reality of star formation suppression in the bar structure from multiple angles.

The structure of this paper is as follows. In Section 2, we provide details of the observations and analysis methods as well as the methods for converting them into molecular gas mass surface density and star formation rate surface density. In Section 3, we present the results obtained from this data, comparing the relationship between molecular gas and star formation in each region, particularly focusing on SFE and SFE$_{\mathrm{dense}}$. In Section 4, we discuss the physical mechanisms behind the observed SFE suppression, based on CO linewidths and the balance between turbulent pressure and equilibrium pressure within the molecular gas. Finally, Section 5 is dedicated to the summary and conclusions of this study.

\section{Data and Analysis}
\subsection{ALMA Observations of HCN and $^{13}$CO($J=1-0$)}
\begin{figure}
 \begin{center}
  \includegraphics[width=0.9\linewidth]{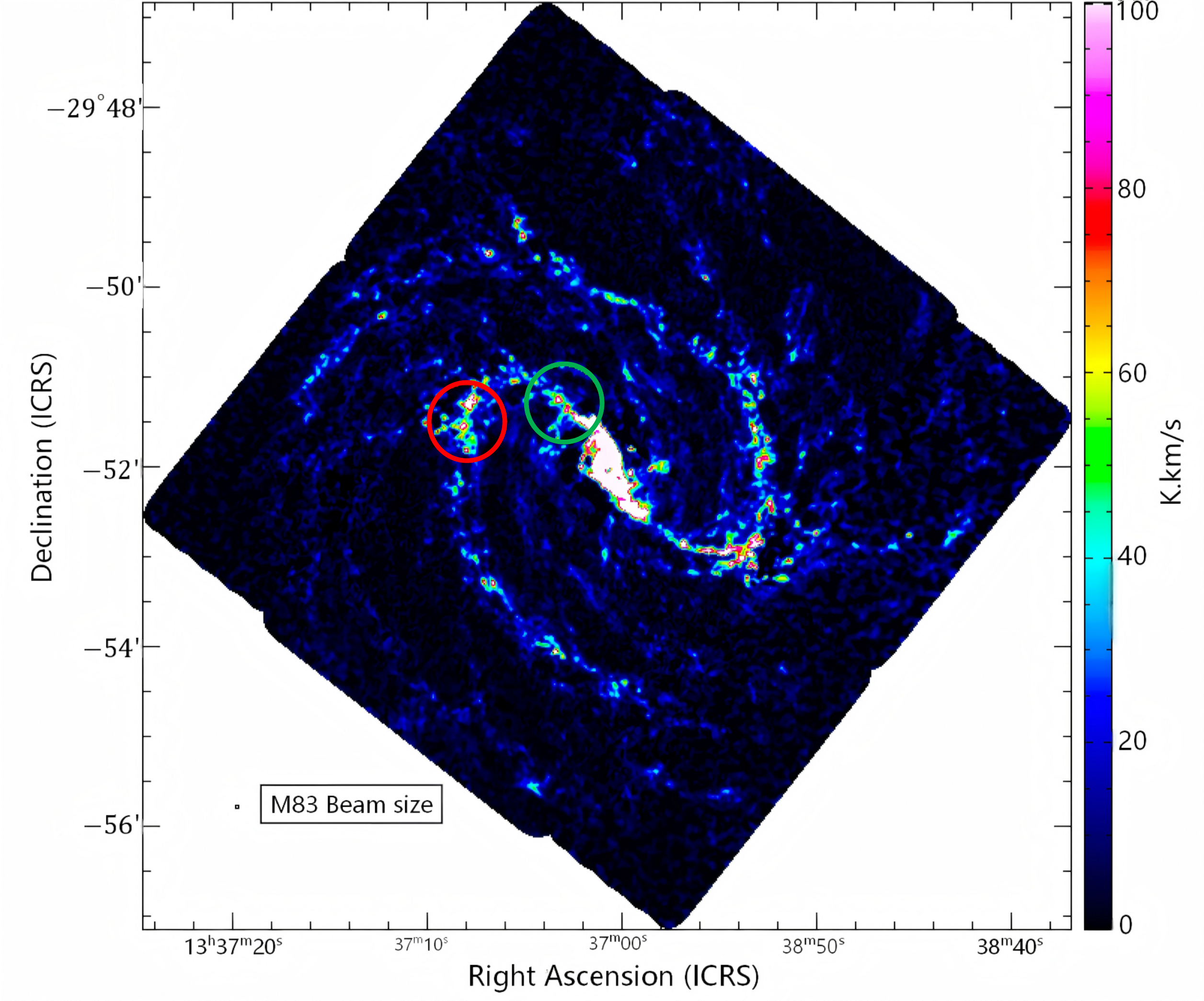}
 \end{center}
\caption{The $^{12}$CO(J=2-1) integrated intensity map of M83 by the PHANGS-ALMA project \citep{Leroy2021}. The green and red circles are the bar and spiral arm regions, respectively, targeted in this work. \\
Alt text: A color map showing the distribution of molecular gas across the face of the M83 galaxy. Green circles mark the bar region, and red circles mark the spiral arms region.}
\label{fig:target_regions}
\end{figure}

We observed HCN ($J=1-0$) and $^{13}$CO ($J = 1-0$) lines in a bar region and a spiral-arm region of M83 (Figure~\ref{fig:target_regions}) with ALMA on May 2015 as a part of the cycle~2 project (ADS/JAO.ALMA \# 2013.1.00889.S).  The phase centers were ($\alpha_{\rm J2000}$, $\delta_{\rm J2000}$) = ($13^{\rm h}37^{\rm m}03.0^{\rm s}$, $-29^{\circ}51'17.3''$) and ($\alpha_{\rm J2000}$, $\delta_{\rm J2000}$) = ($13^{\rm h}37^{\rm m}08.0^{\rm s}$, $-29^{\circ}51'29.8''$) for the bar and the arm, respectively.  The bandwidth and frequency resolution before channel smoothing was 1875~MHz and 0.97~MHz, respectively.  We used J1337-1257 as a bandpass calibrator, J1427-421 and Calisto as flux calibrators, and J1342-2900 as a gain and phase calibrator.  The absolute flux uncertainty was assured to be about 5\%.

The visibility calibrations were carried out by applying the data reduction script provided by the ALMA observatory with the Common Astronomy Software Application (CASA) package \citep{casa2022} version 4.2.2 which contained the pipeline for the Cycle~2 ALMA observation.  Before imaging the molecular lines, we subtracted continuum emission from the visibility data by using line-free channels.  Images were obtained by the CLEAN algorithm with Brigg's weight of 0.5.  In the imaging procedure, the velocity resolution was smooth to be 10~km~s$^{-1}$.  The synthesized beams were $1.83 '' \times 1.58''$ (PA = $-72.2^{\circ}$) and $1.59 '' \times 1.28''$ (PA = $-83.1^{\circ}$) for HCN ($J=1-0$) and $^{13}$CO ($J = 1-0$), respectively.

\subsection{PHANGS-ALMA $^{12}$CO($J=2-1$) Data}
We made use of $^{12}$CO($J=2-1$) line data observed by the ALMA large program PHANGS-ALMA \citep{Leroy2021}. The $^{12}$CO($J=2-1$) data of PHANGS-ALMA were obtained by combining data from ALMA 12m array and the Morita Atacama Compact Array (ACA; composed of the 7m array and four 12m total power telescopes). This enables high-sensitivity imaging with minimal short-spacing loss to capture both extended and compact structures. The typical beam size of the data is about 1~arcsec, and the noise level is about 85~mK at a velocity resolution of 2.54~$\mathrm{km~s^{-1}}$.

\subsection{H$\alpha$ Data}
We used the H$\alpha$ line data observed with the CTIO 1.5m telescope as a part of SINGG project \citep{Meurer2006} to estimate the surface density of SFR. The H$\alpha$ data were obtained using the 6568/28 narrow-band filter, after subtracting the continuum emission with the R-band data. The median FWHM seeing of the data was about 1.6~arcsec.

\subsection{Angular Resolution and Images}
In the following analyses, we use the data smoothed to a common spatial resolution of 9~arcsec, corresponding to 200 pc at the distance of M83 to ensure consistency.

In nearby galaxies, it is known that the SFR strongly correlates with the molecular gas mass on spatial scales of approximately 200 pc \citep{Kennicutt2012}.  In addition, this scale has been shown to be a suitable scale for evaluating dynamical effects like pressure within galactic disks \citep{Sun2018} and is also effective for comparing different galactic environments such as the bar and spiral-arm regions \citep{Querejeta2021}. Large surveys like the PHANGS-ALMA project commonly acquire data at $\sim$100~pc scales \citep{Leroy2021}. On the other hand, at higher spatial resolutions below 100 pc, the correlation between the molecular gas mass and the SFR is known to breakdown as individual star-forming regions are spatially resolved \citep{Onodera2010, Schruba2010}.  The angular resolutions of the ALMA data used in this study are less than 1.6 arcsec, corresponding to approximately 35 pc. Therefore, we smoothed the data to a resolution of 9 arcsec.

Recent HCN observations of NGC~4321 at 260~pc resolution also have demonstrated that this scale is suitable for distinguishing different environments such as the galactic center, bar, and arms, and for examining in detail the relationship between dense gas and star formation in each region \citep{Neumann2024}. In addition, by using the spatial scale of 200 pc, we can directly compare our results with the results of NGC 4321.

\begin{figure*}
 \begin{center}
  \includegraphics[width=0.85\linewidth]{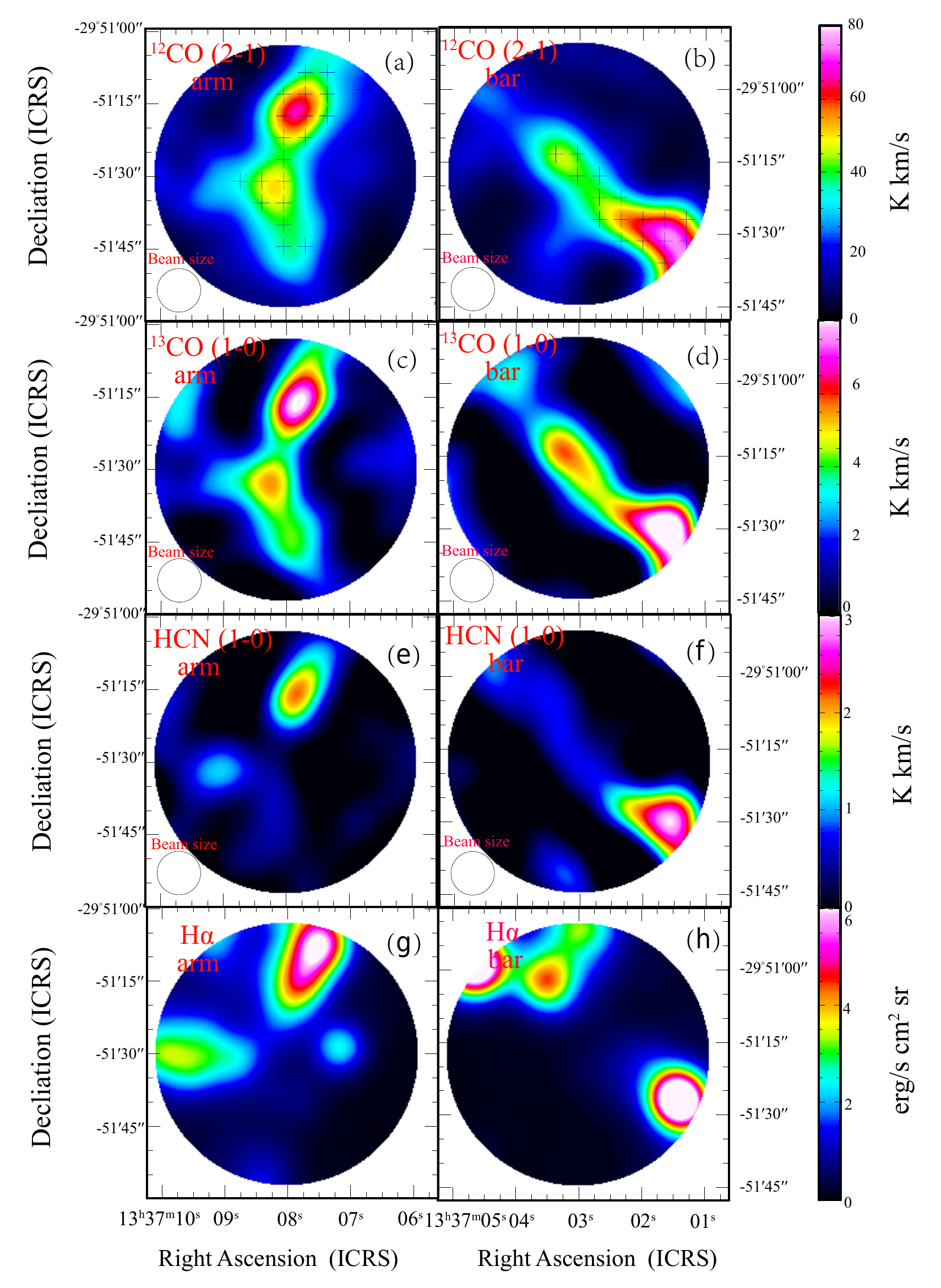}
 \end{center}
\caption{Integrated intensity maps of $^{12}$CO($J=2-1$) in (a) the arm and (b) the bar, $^{13}$CO($J=1-0$) in (c) the arm and (d) the bar, and HCN($J = 1-0$) in (e) the arm and (f) the bar, and H$\alpha$ in (g) the arm and (h) the bar at a common spatial resolution of 9~arcsec (approximately 200~pc). The cross marks marks indicate the data points analyzed in this study. \\
Alt text: An eight-panel grid of intensity maps. Each row compares spiral arm (left) and bar (right) regions, showing the distribution of different molecular gas tracers and H$\alpha$ emission at a common spatial scale.}
\label{fig:integrated_intensity}
\end{figure*}

Figure~\ref{fig:integrated_intensity} shows the integrated intensity maps of the molecular gas emission lines and the H$\alpha$ emission on a scale of 200 pc for the spiral arm and bar structure of M83. These images were obtained by smoothing the original data with a Gaussian kernel using the CASA task imsmooth.

\subsection{Derivation of Integrated Intensity}
At positions where both $^{13}$CO($J=1-0$) and H$\alpha$ emissions are detected with a signal-to-noise ratio (S/N) of 5$\sigma$ or greater, the integrated intensities of $^{12}$CO($J=2-1$), $^{13}$CO($J=1-0$), and HCN($J = 1-0$) were calculated as follows in order to reduce noise contamination to the values.

Due to the primary beam corrections of the ALMA observations, the noise level in the molecular line data increases toward the edge of the field of view. Therefore, we need to evaluate the RMS noise ($\sigma_{\text{rms}}$) in line-free velocity channels from position to position. The final signal integration range was determined by identifying channels with signals exceeding 3$\sigma_{\text{rms}}$ consecutively and including adjacent channels with signals above 1.5$\sigma_{\text{rms}}$. Then, the velocity-integrated intensities ($I$) at each position were calculated using the formula:
\begin{equation}
    I = \sum_i T_{B,i} \times \Delta v \ ({\rm K~km~s^{-1}})
\end{equation}
where $T_{B,i}$ is the brightness temperature and $\Delta v$ is the velocity channel width.

The error in the integrated intensity ($\Delta I$) for each positions were estimated. This error was calculated based on the noise characteristics of the spectrum using the formula:
\begin{equation}
    \Delta I = \sigma_{\text{rms}} \times \Delta v \times \sqrt{N} \ ({\rm K~km~s^{-1}})
\end{equation}
where $\sigma_{\text{rms}}$ is the RMS noise in the line-free channels at the position, $\Delta v$ is the velocity channel width, and $N$ is the number of channels in the integration range.

\subsection{Molecular Gas Surface Density}
The molecular gas surface density ($\Sigma_{\text{mol}}$($^{12}$CO)) was derived from the integrated intensity of $I_{\text{CO(2-1)}}$ using the following formula:
\begin{equation}
\Sigma_{\mathrm{mol}}(\mathrm{^{12}CO}) = \frac{\alpha_{\rm CO}}{R_{21}}I_{\rm CO(2-1)}\cos i \ ({\rm M_{\odot}pc^{-2}})
\end{equation}
where $\alpha_{\text{CO}}$ is the CO-to-H$_2$ conversion factor with units of $\mathrm{M_\odot\ pc^{-2}\ (K\ km\ s^{-1})^{-1}}$, $I_{\text{CO($J = 2-1$)}}$ is the velocity-integrated intensity of CO($J = 2-1$) in units of $\mathrm{K\ km\ s^{-1}}$, $R_{21}$ is the integrated intensity ratio between $^{12}$CO($J=2-1$) and $^{12}$CO($J = 1-0$) ($R_{21} = I_{\text{CO}(2-1)} / I_{\text{CO}(1-0)}$), and $i$ is the inclination angle of the galaxy.

For the $\alpha_{\mathrm{CO}}$, we adopted the value given by \citet{Lee2024} for M83. In this study, the bar and spiral arm regions are located at approximately 1.0 kpc and 2.1 kpc, respectively, from the center of M83. Because both positions fall within the 1–3 kpc radial bin defined by \citet{Lee2024}, we employed $\alpha_{\mathrm{CO}} = 1.82\ \mathrm{M_\odot\ pc^{-2}\ (K\ km\ s^{-1})^{-1}}$ for both positions.

Spatial variations of the $R_{21}$ across galactic structures have been reported in M83, with typical values of $\sim$0.8--0.9 in the bar and $\sim$0.6--0.7 in the spiral arms \citep{Koda2020, Koda2025}. We adopted $R_{21}$ = 0.65 and 0.85 for the arm and bar regions, respectively.

From the $^{13}$CO($J=1-0$) data, we directly estimated the column density of $^{13}$CO under the assumption of local thermodynamic equilibrium (LTE) with the optically thin condition. In this calculation, we assumed a constant excitation temperature of $T_{\rm ex} = 20$~K. This value is commonly adopted for molecular gas in nearby spiral galaxies and corresponds to moderate excitation conditions. Multi-line CO analyses of nearby galaxies, including M83, indicate that the bulk molecular gas traced by low-$J$ CO transitions typically resides in cold to moderately excited phases with characteristic temperatures of $\sim$10--30~K, with effective excitation temperatures often falling in the range of $\sim$15--25~K \citep{Leroy2022}.

Under the assumptions, the column density of the $^{13}$CO ($N(^{13}\mathrm{CO})$) was calculated from the integrated intensity of $^{13}$CO($J=1-0$) $I_{\rm ^{13}CO(1-0)}$ as following \citet{Garden1991}:
\begin{equation}
N(^{13}\mathrm{CO}) = \frac{3kU(T_{\rm ex})}{8\pi^3S\mu^2\nu}\exp\left(\frac{E_{\rm u}}{kT_{\rm ex}}\right)\times I_{\rm ^{13}CO(1-0)}\times \cos i \ ({\rm cm^{-2}}),
\end{equation}
where k is the Boltzmann constant, $U(T_{\mathrm{ex}})$ is the partition function, $T_{\rm ex}$ is the excitation temperature of $^{13}$CO($J=1-0$), S is the line intensity of the transition, $\mu$ is the dipole moment, $\nu$ is the frequency of the transition,  $E_{\rm u}$ is the upper state energy of the transition, and $I_{\rm ^{13}CO(1-0)}$ is the integrated intensity of $^{13}$CO($J=1-0$). We used the line parameters of $^{13}$CO($J=1-0$) listed in the Cologne Database for Molecular Spectroscopy (CDMS) \citep{Muller2001}. The parameters are summarized in Table 2.

\begin{table}
  \tbl{Line parameters of $^{13}$CO($J=1-0$).}{%
  \begin{tabular}{p{3cm}c}  
      \hline
      Parameters & Value \\
      \hline
      $T_{\rm ex}$ & 20~K \\
      $U(T_{\mathrm{ex}})$ & 15.8152 \\
      $S\mu^2$ & $2.4393$  \\
      $\nu$ & $110.201354\ \mathrm{GHz}$ \\
      $E_{\rm u}$ & 5.2898~K \\
      \hline
    \end{tabular}}
    \label{tab:co13_parameters}
\begin{tabnote}
All parameters are referred from CDMS . 
\end{tabnote}
\end{table}

From the $^{13}$CO($J=1-0$) column density, we calculated the molecular gas surface density($\Sigma_{\rm mol}({\rm ^{13}CO})$). The H$_2$ column density was calculated by assuming an isotopic ratio of $^{12}$C/$^{13}$C = 57 \citep{Langer1990}, and the C$^{18}$O fractional abundance of N(C$^{18}$O)/N(H$_2$) = $1.7 \times 10^{-7}$ \citep{Goldsmith1997} with the oxygen isotope ratio of $^{16}$O/$^{18}$O = 560 \citep{Wilson1999}. We converted the surface density of H$_2$ to that of molecular gas by including a contribution of helium (factor of 1.36). Combining these conversions, the approximate relation from $I_{\rm ^{13}CO(1-0)}$ to molecular gas surface density is:
\begin{equation}
    \Sigma_{\text{mol}}(^{13}\mathrm{CO}) \approx 17.2 \times I_{^{13}\mathrm{CO}(1-0)} \times \cos i \ ({\rm M_{\odot}~pc^{-2}}).
\end{equation}

We note that adopting a lower excitation temperature of $T_{\text{ex}}$ = 15~k would reduce the derived $^{13}$CO-based gas surface density by $\sim$20\%, while adopting a higher value of $T_{\text{ex}}$ = 25~k) would increase it by $\sim$15\%, primarily due to the change in the partition function. A $\pm$ 5\% change in the excitation temperature do not fully remove the discrepancy between $\Sigma_{\rm mol}(^{12}\mathrm{CO})$ and $\Sigma_{\rm mol}(^{13}\mathrm{CO})$ discussed in Section~3.2.

\subsection{SFR Surface Density}
The surface density of star formation rate ($\Sigma_{\text{SFR}}$) is defined as the stellar mass formed per unit time and unit area, and is a key parameter for quantifying the spatial distribution of star formation activity in galaxies. In this study, $\Sigma_{\text{SFR}}$ was calculated from the extinction-corrected H$\alpha$ line intensity $I_{\rm H\alpha}$ \citep{Meurer2006}. The extinction correction was applied assuming a constant Balmer extinction coefficient ($A_{\text{H}\alpha}$). In this study, we used $A_{\text{H}\alpha} = 0.24$~mag \citep{Dopita2010}. The $I_{\rm H\alpha}$ value in the unit of $\mathrm{erg\ s^{-1}\ cm^{-2}\ sr^{-1}}$ was converted to $\Sigma_{\mathrm{SFR}}$ using the conversion factor provided in \citet{Kennicutt2012} as 
\begin{equation}
\Sigma_{\mathrm{SFR}} = 6.3 \times 10^{2} \times I_{\mathrm{H}\alpha} \times\cos i \ ({\rm M_{\odot}~pc^{-2}~yr^{-1}}).
\end{equation}

\subsection{HCN and Dense Gas}
Star formation is thought to occur in high-density cores with $n > 10^4$~cm$^{-3}$ within molecular clouds. To detect this dense gas, we used the HCN($J = 1-0$) emission line. Due to its high critical density ($\sim 10^5$~cm$^{-3}$), HCN is thought to trace the dense gas cores that are the material for star formation. The dense gas surface density ($\Sigma_{\text{dense}}$) was estimated from the HCN integrated intensity ($I_{\text{HCN}}$) using a conversion factor given by \citet{Neumann2024} as
\begin{equation}
    \Sigma_{\mathrm{dense}}= 10 \times I_{\mathrm{HCN}} \times \cos i~({\rm M_{\odot}~pc^{-2}}).
\end{equation}

\section{Results}
In this section, we compare the surface densities of molecular gas $\Sigma_{\rm mol}({\rm ^{12}CO})$, $\Sigma_{\rm mol}({\rm ^{13}CO})$ and $\Sigma_{\rm dense}$ and the surface density of star formation rate ($\Sigma_{\rm SFR}$) in the bar and spiral arm regions of M83, and clarify the environmental effects on SFE and SFE$_{\mathrm{dense}}$.

\subsection{Correlation between $\Sigma_{\text{SFR}}$ and $\Sigma_{\rm mol}({\rm ^{12}CO})$}
\begin{figure*}
 \begin{center}
  \includegraphics[width=\linewidth]{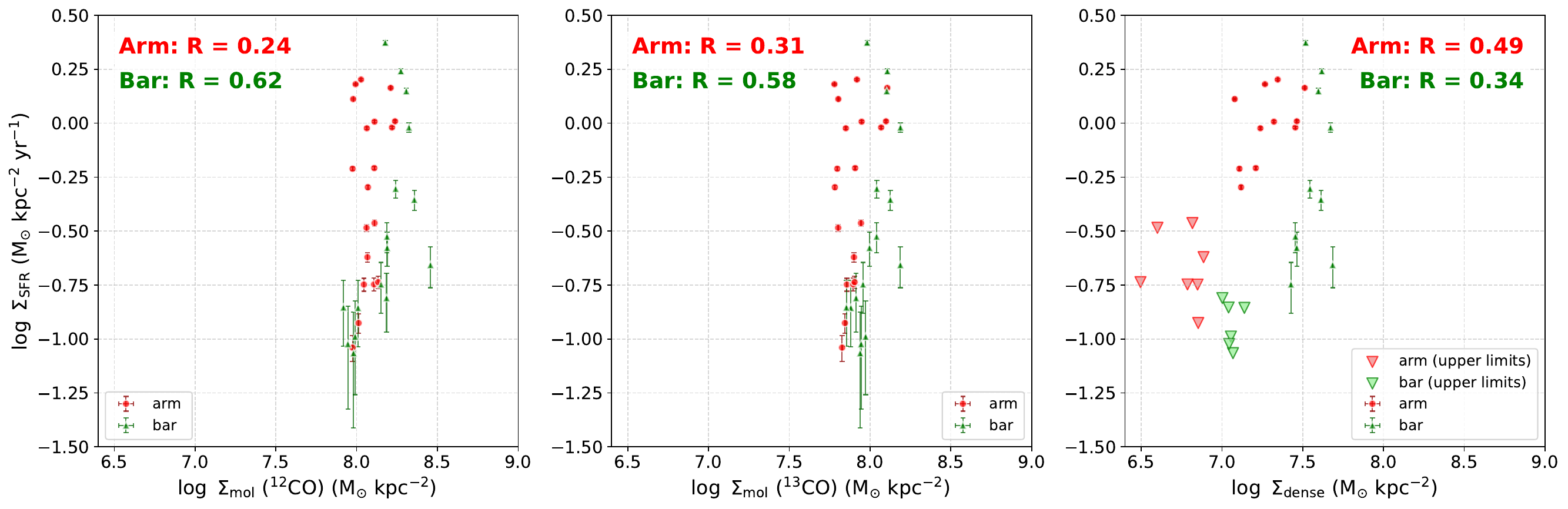}
 \end{center}
\caption{Correlation plots between $\Sigma_{\mathrm{SFR}}$ and molecular gas surface densities in M83. 
(a) $\Sigma_{\mathrm{SFR}}$ versus $\Sigma_{\rm mol}({\rm ^{12}CO})$ (b) $\Sigma_{\mathrm{SFR}}$ versus $\Sigma_{\rm mol}({\rm ^{13}CO})$, and 
(c) $\Sigma_{\mathrm{SFR}}$ versus $\Sigma_{\mathrm{dense}}$. 
The bar regions are shown in green, and the spiral arm regions in red. 
Triangle symbols indicate data points with upper limits on $\Sigma_{\mathrm{dense}}$. 
R denotes the Pearson correlation coefficient. \\
Alt text: Three scatter plots arranged horizontally. Each plot shows data points with green for bar regions and red for spiral arm regions, plotting SFR density against different molecular gas density tracers. Triangles denote upper limit measurements.}
\label{fig:sfe_relations}
\end{figure*}

Figure~\ref{fig:sfe_relations}a shows the relationship between $\Sigma_{\rm mol}({\rm ^{12}CO})$ and $\Sigma_{\mathrm{SFR}}$ in the bar and spiral arm regions.  $\Sigma_{\rm mol}({\rm ^{12}CO})$ in the bar region was observed to be systematically about 1.3 times higher than those in the spiral arm region. However, no significant systematic difference can be seen in the $\Sigma_{\rm SFR}$ value between the two regions. This result suggests that the SFE($^{12}$CO) which is the ratio of $\Sigma_{\text{SFR}} / \Sigma_{\rm mol}({\rm ^{12}CO})$ is relatively lower in the bar region than the spiral arm, although several positions within the bar exhibit SFE($^{12}\mathrm{CO}$) values comparable to those in the arms.  In fact, the average SFE($^{12}\mathrm{CO}$) in the bar is smaller than that in the arm by a factor of 0.65 (Table \ref{tab:SFE_average}).

\begin{table*}[h]
\centering
\caption{Average values of the SFE and $\sigma_v$ in the arm and bar regions}
\label{tab:SFE_average}

\begin{tabular}{l c c c c c}
\toprule
 & Average SFE ($^{12}\text{CO}$)$^\dagger$ & Average SFE ($^{13}\text{CO}$) & Average SFE$_{\text{dense}}$ & \multicolumn{2}{c}{Average $\sigma_v$} \\
\cmidrule(l){2-2} \cmidrule(l){3-3} \cmidrule(l){4-4} \cmidrule(lr){5-6}
 & ($\times 10^{-8}~\text{yr}^{-1}$) & ($\times 10^{-8}~\text{yr}^{-1}$) & ($\times 10^{-8}~\text{yr}^{-1}$) & $^{12}\text{CO}$ (km/s) & $^{13}\text{CO}$ (km/s) \\
\midrule
Arm & $0.56 \pm 0.02$ & $0.87 \pm 0.01$ & $5.04 \pm 0.06$ & 7.91 & 4.2 \\
Bar & $0.36 \pm 0.07$ & $0.51 \pm 0.01$ & $1.80 \pm 0.07$ & 12.43 & 6.3 \\
Ratio (Bar/Arm) & $0.65 \pm 0.01$ & $0.58 \pm 0.01$ & $0.35 \pm 0.01$ & 1.57 & 1.50 \\
\bottomrule
\end{tabular}
\vspace{2pt}
\parbox{0.9\linewidth}{\footnotesize
$^\dagger$ $R_{21}=0.65$ for the arm region, $R_{21}=0.85$ for the bar region.
}
\end{table*}

\subsection{Compare between $\Sigma_{\text{SFR}}$ and $\Sigma_{\rm mol}({\rm ^{13}CO})$}
Figure~\ref{fig:sfe_relations}b shows the correlation plots between $\Sigma_{\rm mol}({\rm ^{13}CO})$ data and $\Sigma_{\text{SFR}}$. Compared with the $\Sigma_{\rm mol}({\rm ^{12}CO})$ values, the $\Sigma_{\rm mol}({\rm ^{13}CO})$ values are systematically lower in both the arm and bar regions (also see Figure~\ref{fig:parameter_comparison}c in Appendix 1).  As a result, the SFE based on $\Sigma_{\rm mol}({\rm ^{13}CO})$, SFE($^{13}\mathrm{CO}$)=$\Sigma_{\text{SFR}}$ / $\Sigma_{\rm mol}({\rm ^{13}CO})$, is higher than SFE($^{12}\mathrm{CO}$). The discrepancy between $\Sigma_{\rm mol}({\rm ^{12}CO})$ and $\Sigma_{\rm mol}({\rm ^{13}CO})$ is likely attributable to the assumed $T_{\text{ex}}$, the fractional abundance of C$^{18}$O, and the isotopic ratios of $^{16}$O/$^{18}$O and $^{12}$C/$^{13}$C.  In this analysis, we assumed $T_{\text{ex}}$ =20 K. If $T_{\text{ex}}$ = 25 K is adopted, the derived $\Sigma_{\rm mol}({\rm ^{13}CO})$ increases by $\sim$15\%.         

Despite this systematic offset in the absolute values of the molecular gas surface density, the relationship between $\Sigma_{\rm mol}({\rm ^{13}CO})$ and $\Sigma_{\text{SFR}}$ revealed by $^{13}$CO($J=1-0$) is qualitatively similar to the trend seen with $^{12}$CO($J=2-1$). In fact, the ratio of the average SFE($^{13}\mathrm{CO}$) in the bar to that in the arm (0.58) remains consistent with the ratio of SFE($^{12}\mathrm{CO}$). The average values of SFE($^{13}\mathrm{CO}$) and the ratio are summarized in Table \ref{tab:SFE_average}. 

\subsection{Relationship between Dense Gas and Star Formation Rate}
To clarify the relationship between dense gas and star formation activity, Figure~\ref{fig:sfe_relations}c shows the correlation plots between $\Sigma_{\text{dense}}$ and $\Sigma_{\text{SFR}}$. Similar to $\Sigma_{\text{mol}}(^{12}\text{CO})$ and $\Sigma_{\text{mol}}(^{13}\text{CO})$, $\Sigma_{\text{dense}}$ exhibits a tendency to be higher in the bar region than in the spiral arm region. The dense gas star formation efficiency ($\text{SFE}_{\text{dense}} \equiv \Sigma_{\rm SFR} / \Sigma_{\text{dense}}$) is on average about 0.35 times lower in the bar region than that in the spiral arm region. These average values are summarized in Table \ref{tab:SFE_average}. Given that $\Sigma_{\text{dense}}$ is smaller than $\Sigma_{\text{mol}}(^{12}\text{CO})$ and $\Sigma_{\text{mol}}(^{13}\text{CO})$, the corresponding values of $\text{SFE}_{\text{dense}}$ are approximately 4-10 times higher than those of $\text{SFE}(^{12}\text{CO})$ and $\text{SFE}(^{13}\text{CO})$.

\section{Discussion}
In this section, we discuss the physical mechanisms responsible for the observed SFE suppression focusing on dense gas, molecular gas dynamics, and environmental conditions such as the dynamical equilibrium pressure and the turbulent pressure.

\subsection{Suppression of SFE in the Bar Region}
From the comparisons, we found that the SFE in the bar region of M83 is significantly suppressed in both moderate density gas traced by CO and dense gas traced by HCN compared with the spiral arm. We summarize our findings to discuss the implied physical picture in the following subsections.

It is noteworthy that the results based on $^{12}$CO($J=2-1$) and $^{13}$CO($J=1-0$) are qualitatively consistent with each other. $^{12}$CO($J=2-1$) is optically thick and traces the large-scale distribution of molecular gas, while $^{13}$CO($J=1-0$) is relatively optically thin and more directly reflects the column density of molecular gas \citep{Pineda2008}. The consistent results obtained from these two tracers support the conclusion that the observed SFE suppression in the bar region is a real physical phenomena and not an artifact caused by the properties of the selected molecular gas tracers.

We also confirmed a decrease in the SFE$_{\mathrm{dense}}$ within the bar region. These results indicate that the impact of the bar structure is not limited to the accumulation phase of molecular gas, but also affects the phase from the dense gas to the star formation. Previous studies have shown that the relationship between dense gas and star formation can vary with environment \citep{Usero2015, Bigiel2016}, and such environmental dependence of SFE$_{\mathrm{dense}}$ has also been reported in recent observations of NGC~4321 \citep{Neumann2024}. \citet{Neumann2024} discussed that the SFE$_{\mathrm{dense}}$ suppression in the bar could be due to that the gas-to-star conversion process itself become inefficient or that HCN traces a more widespread molecular gas similar to that traced by CO. Thus, our results suggest that, in addition to the suppression of molecular gas accumulation \citep{Muraoka2016}, the gas-to-star evolutionary process itself becomes inefficient under the dynamical conditions in the bar of M83. 

We note that \citet{Neumann2024} also suggested that the HCN emission in the bar region might trace a more widespread molecular gas rather than the dense molecular gas associated with star formation due to subthermally excited condition of HCN. If this is the case in M83, the observed suppression of SFE$_{\mathrm{dense}}$ could partially reflect a change in the properties of tracer rather than a change in the SFE of dense gas. However, even under this interpretation, there is no contradiction with the conclusion drawn from the analyses of SFE($^{12}$CO) and SFE($^{13}$CO) that the SFE is suppressed in the bar. Therefore, even if HCN traces a somewhat more diffuse component, our results support that the global process of converting the available molecular reservoir into stars is suppressed. To decisively verify whether the conversion from dense gas ($n > 10^4-10^5~\mathrm{cm^{-3}}$) to stars is suppressed, additional observations of dense-gas tracers such as N$_2$H$^+$ are essential. This is the future work. Nevertheless, the most straightforward interpretation of our multi-tracer analysis is that dynamical effects in the bar region, such as strong shear and fast cloud-cloud collisions, directly hinder the gravitational collapse and growth of dense structures \citep{Tubbs1982, Fujimoto2014a}, leading to the observed suppression of star formation efficiency across different gas phases. 

The impact of galactic dynamics on the star formation activity has also been suggested in the disk regions, in addition to the bar region. For example, \citet{Koda2025} found systematic variation of the $^{12}$CO($J = 2-1$)/$^{12}$CO($J = 1-0)$ line ratios in M83 and discussed that these variations indicate that the physical conditions of molecular gas, such as density and excitation state, evolve in a manner that is structured by the underlying galactic potential and synchronized with galactic rotation. Such results suggest that large-scale dynamical processes regulate the molecular gas properties across the disk, beyond the influence of local star formation feedback.

\subsection{CO Line Width and SFE}
One plausible origin of the SFE suppression is the strength of non-circular motions of molecular gas in the bar region. Previous studies have suggested that dynamical effects induced by non-circular motions, such as strong shocks, large shears, and fast cloud-cloud collisions, may suppress star formation \citep[e.g.,][]{Tubbs1982, Fujimoto2014a, Fujimoto2020, Maeda2023, Koda2025}. Furthermore, numerous observational studies of CO reported that the linewidth of CO emission lines is larger in the bar regions than in the arm regions, which supports the effects of non-circular motions in bars \citep[e.g.,][]{Watanabe2011, Muraoka2016, Maeda2018}. Here, we focus on the detail relationship between CO emission linewidth and SFE values in regions including the bar and spiral arms.

The CO line width ($\sigma_{\text{v}}$) was evaluated by the moment 2 value as:
\begin{equation}
\sigma_{\mathrm{v}} = \sqrt{\frac{\sum_i T_{B,i} (v_i - \bar{v})^2  \Delta v}{\sum_i T_{B,i}  \Delta v}} \quad (\mathrm{kms^{-1}}).
\end{equation}
Here, $v_i$ is the velocity of the $i$-th channel, and $\bar{v}$ is the intensity-weighted mean velocity of the line.
This intensity-weighted averaging method has the advantage of preserving information about the gas distribution on molecular cloud scales \citep{Leroy2016}. Using the CO intensity as a statistical weight prevents the dilution of the average value by regions where CO is not detected.

\begin{figure*}
 \begin{center}
  \includegraphics[width=0.8\linewidth]{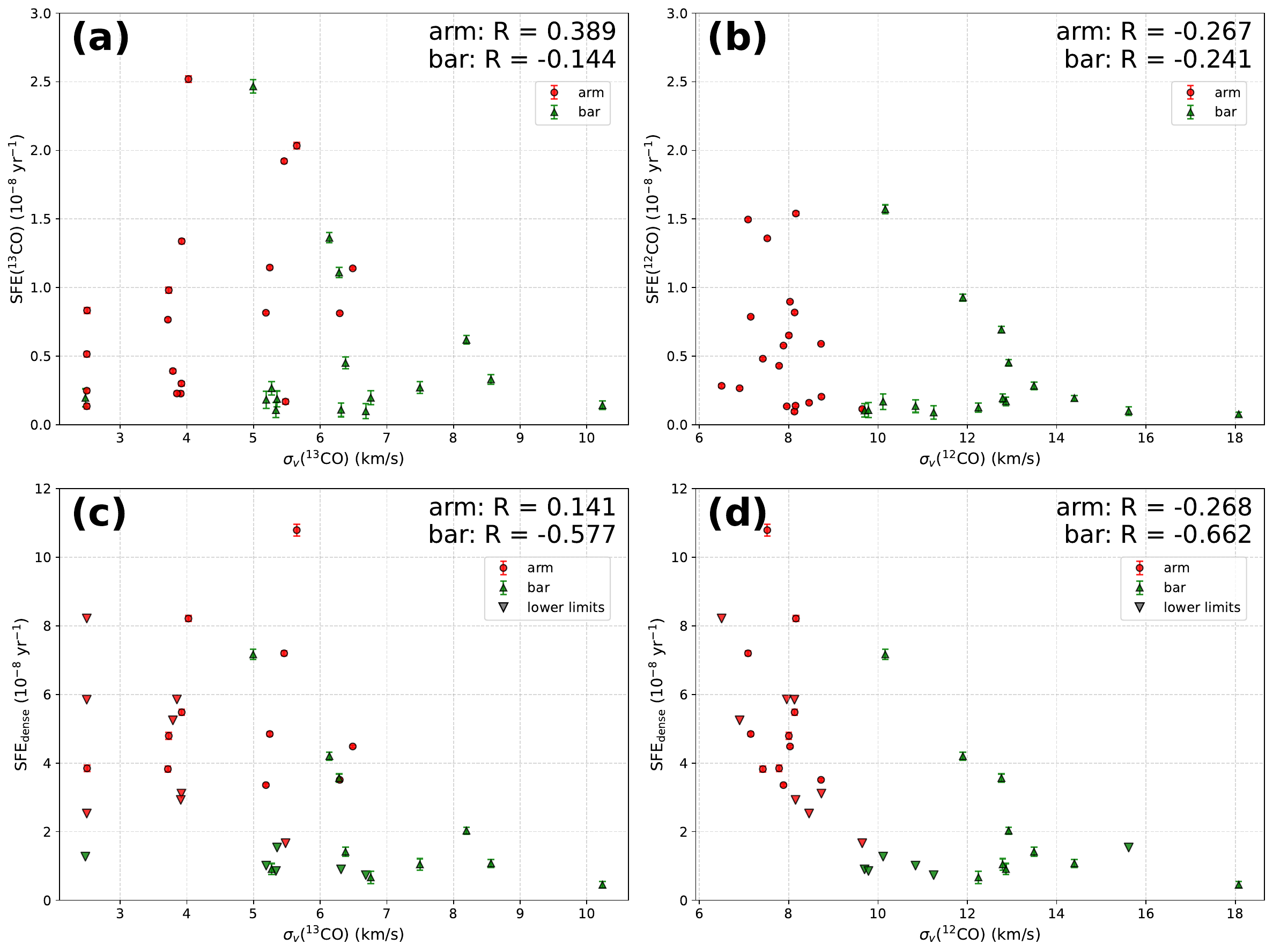}
 \end{center}
\caption{Relationship between CO linewidth and star formation efficiency in M83. (a) $\sigma_v(^{13}\mathrm{CO})$ versus SFE($^{13}$CO), (b) $\sigma_v(^{12}\mathrm{CO})$ versus SFE($^{12}$CO). (c) $\sigma_v(^{13}\mathrm{CO})$ versus SFE$_{\mathrm{dense}}$ and (d) $\sigma_v(^{12}\mathrm{CO})$ versus SFE$_{\mathrm{dense}}$. The bar, spiral arm are indicated in green and red, respectively. The inverted triangular symbols show lower limits on SFE$_{\mathrm{dense}}$. R indicates the value of the Pearson correlation coefficient. A negative correlation is seen in all plots, indicating that the SFE is systematically lower in regions with broader line widths. \\
Alt text: A four-panel grid of scatter plots. The left column plots $^{13}$CO($J=1-0$) line width against two star formation efficiency measures, and the right column plots $^{12}$CO($J=2-1$) line width against the same two efficiency measures. Red points represent arm regions, and green points represent bar regions.}
\label{fig:velocity_width}
\end{figure*}

Figure~\ref{fig:velocity_width} shows the relationship between the linewidth of $^{13}$CO($J=1-0$)($\sigma_v(^{13}\mathrm{CO})$) and $^{12}$CO($J=2-1$)($\sigma_v(^{12}\mathrm{CO})$), and the SFE values in the bar and arm regions. A clear negative correlation is found between the linewidth and the SFE values, including SFE($^{12}$CO), SFE($^{13}$CO), and SFE$_{\mathrm{dense}}$. This trend is consistent for both tracers, aligning with the results reported by \citet{Maeda2023}. Furthermore, the linewidth values for both $^{12}$CO($J=2-1$) and $^{13}$CO($J=1-0$) are systematically larger in the bar region than in the arm region, with the average linewidth in the bar being approximately 1.5 times greater. Regions of lower SFE consistently correspond to regions of larger linewidth. Even within the bar region, the SFE varies by more than a factor of five, and regions with lower SFE tend to have larger line widths. This overall negative correlation strongly supports the scenario in which greater non-circular motion, manifested as broader linewidths, leads to a reduction in the SFE.

\citet{Teng2023} reported that the optical depth of the $^{12}$CO emission decreases in regions with broader linewidths, which results in a smaller $\alpha_{\mathrm{CO}}$. Because we adopted a constant $\alpha_{\mathrm{CO}}$ in this analysis, the $\Sigma_{\mathrm{mol}}(^{12}\mathrm{CO})$ may be overestimated if the $\alpha_{\mathrm{CO}}$ decreases with the linewidth. In such cases, the SFE($^{12}$CO) would be underestimated in the regions. On the other hand, the $\Sigma_{\mathrm{mol}}(^{13}\mathrm{CO})$, which is generally optically thinner, is expected to be less affected by this effect. Therefore, SFE($^{13}$CO) provides a more reliable estimate of the SFE, and lower SFE($^{13}$CO) in the bar indicates that the SFE is intrinsically lower in the region.

As indicated by the large linewidths, strong shocks, large shears, and fast cloud-cloud collisions are expected in the bar region. The combined action of these effects is thought to reduce star formation efficiency. From the numerical simulations, strong shocks and large shears associated with non-circular gas flows induced by the gravitational potential of the bar are considered to suppress star formation by destroying molecular clouds and inhibiting the formation of the clouds themselves \citep[e.g.,][]{Tubbs1982, Renaud2015}. Sub-parsec scale simulations showed that faster collision speeds shorten the gas accretion phase of the formed cloud cores, suppressing core growth and massive star formation\citep{Takahira2014, Takahira2018}. Furthermore, \citet{Fujimoto2020} suggested that cloud-cloud collision velocities in the bar region are larger than in other regions, likely because cloud motions are perturbed into elliptical orbits by gravitational interactions between clouds. From these results, they proposed a model where fast collisions in the bar region suppress massive star formation \citep[see also][]{Fujimoto2014a, Fujimoto2014b}.

From Figure \ref{fig:velocity_width}, we confirmed large linewidth of CO lines in the bar region where the SFE values are lower than those in the arm. As reported by \citet{Maeda2021}, we also consider that non-circular motions caused by the gravitational potential of bar are the cause of the fast cloud-cloud collisions. 

Lower SFE$_{\mathrm{dense}}$ is also observed in regions with larger linewidths. If HCN traces the dense gas, this implies that the high-density cores generated in the bar region have their internal mass accretion hindered by the intense turbulent environment, consequently causing them to stagnate in a state of reduced SFE$_{\mathrm{dense}}$. This interpretation is consistent with the conclusion reached by \citet{Neumann2024} for the bar in NGC~4321, that gas dynamics can significantly impact the efficiency with which dense gas is converted into stars. To verify this scenario, further observations using alternative dense gas tracers less affected by excitation conditions, such as N$_2$H$^+$, are necessary as we discussed in Section 4.1.

\subsection{Pressure and Dynamical State}

The large CO line widths observed in the bar region suggest the presence of high-velocity cloud-cloud collisions and strong non-circular motions. Such kinematics not only disrupt individual cloud motions but also inject energy, enhancing the turbulent pressure ($P_\mathrm{turb}$) within the molecular gas. This elevated internal pressure acts against the self-gravitational contraction of molecular clouds, thereby suppressing star formation.

To quantitatively evaluate the dynamical state of the gas, we compare this internal turbulent pressure with the dynamical equilibrium pressure ($P_\mathrm{DE}$). Before introducing the equations, it is essential to clarify the physical setup. Following the framework of \citet{Sun2020}, we model the interstellar medium as a clumpy gas layer embedded within an extended galactic disk under vertical hydrostatic equilibrium.

In this setup, $P_\mathrm{DE}$ is not an external contact force acting on the lateral surface of a cloud. Rather, it represents the total vertical gravitational weight per unit area of the gas layer. To maintain hydrostatic equilibrium and prevent the gas layer from collapsing vertically, the midplane internal pressure must balance this gravitational load. Therefore, while the terms used to evaluate $P_\mathrm{DE}$ describe gravitational forces (and inherently include the gravitational constant $G$), their dimension is force per unit area, making $P_\mathrm{DE}$ physically equivalent to the required equilibrium pressure.

\subsubsection{Estimation of $P_{\rm DE}$ and $P_{\rm turb}$}
In previous studies, $P_{\text{DE}}$ of the interstellar medium (ISM) has often been evaluated by assuming a gas disk smoothed on kpc scales \citep[e.g.,][]{Blitz2004, Blitz2006}. However, the real ISM, especially the molecular gas phase, shows multiple small-scale structure due to turbulence and gravitational instability. Since this structure locally enhances the self-gravity of the gas, the $P_{\text{DE}}$ estimates based on a smooth disk model is thought to be underestimated. 

\citet{Sun2020} developed the method to estimate the $P_{\text{DE}}$ by assuming the interstellar medium in the galactic disk as two components, a thin clumpy layer of molecular gas including molecular clouds and a smooth plane-parallel outer layers sandwiching the molecular gas layer. By adding up the contributions of the stellar, atomic, and molecular components in the two parts, they evaluated $P_{\rm DE}$ at the center of a molecular cloud.  A detailed description is given in Appendix A of \citet{Sun2020}. Instead of relying on identifying individual molecular clouds, they directly evaluated the weight of the molecular gas within its local gravitational potential at each pixel. By combining the weight of the molecular gas with the weight originating from the gravitational potential of the extended atomic gas and stars, they evaluated the $P_{\text{DE}}$ at the cloud scale. This method is robust even when molecular clouds are not fully resolved and can reproducibly characterize the behavior of the entire molecular gas reservoir.

We employ their formula to estimate the $P_{\text{DE}}$ and $P_{\text{turb}}$ in M83. In Table~4, we summarized the physical quantities used to estimate the $P_{\text{DE}}$ and $P_{\text{turb}}$ in this work. These quantities were derived from multi-wavelength observational data as shown in Table \ref{tab:additional_properties}.

At the scale of molecular clouds, the vertical gravitational weight per unit area of the molecular gas ($\overline{W}_{\rm cloud}$) is calculated as the sum of three vertical gravitational weight terms per unit area:

\begin{equation}
\begin{aligned}
\overline{W}_{\rm cloud} & = \overline{W}_{\rm cloud}^{\rm self} + \overline{W}_{\rm cloud}^{\rm ext-mol} + \overline{W}_{\rm cloud}^{\rm star} \\
& = \frac{3\pi}{8} G \Sigma_{\mathrm{mol}}^{2} + \frac{\pi}{2} G \Sigma_{\mathrm{mol}} \Sigma_{\mathrm{mol,1\ kpc}} \\
& \quad + \frac{3\pi}{4} G \rho_{*,1\ \mathrm{kpc}} \Sigma_{\mathrm{mol}} D_{\mathrm{cloud}}.
\end{aligned}
\end{equation}

Here, these terms represent the vertical gravitational weight per unit area contributed by different physical components. The first term $\overline{W}_{\rm cloud}^{\rm self}$ represents the weight from the self-gravity. The second term $\overline{W}_{\rm cloud}^{\rm ext-mol}$ is the contribution of the gravity from the entire spatially extended molecular gas disk (represented by the kpc-scale average surface density $\Sigma_{\text{mol,1kpc}}$). The third term $\overline{W}_{\rm cloud}^{\rm star}$ represents the weight contribution arising from the vertical gradient of the stellar gravitational field across the cloud thickness. Near the disk midplane, the vertical stellar gravitational force increases linearly with height from the midplane; integrating this height-dependent stellar gravitational force over the mass distribution of a cloud with thickness $D_{\rm cloud}$ yields the corresponding stellar contribution to the vertical weight term. Here, $G$, $\Sigma_{\rm mol}$, $\Sigma_{\rm mol, 1 kpc}$, $\rho_{\rm *, 1 kpc}$, and $D_{\rm cloud}$ denote the gravitational constant, the surface density of the molecular cloud, the surface density of the molecular cloud on a scale of 1 kpc, the stellar midplane volume density, and the diameter of the molecular cloud, respectively. As suggested by the last term of Equation (9), the effective thickness of the clumpy molecular layer is represented by the cloud diameter $D_{\rm cloud}$. Similarly, the effective thickness of the atomic gas layer is dynamically determined by its vertical velocity dispersion $\sigma_{\rm atom}$ (introduced in Equation 10 below) within the galactic gravitational potential. For simplicity, $\Sigma_{\rm mol}({\rm ^{12}CO})$ or $\Sigma_{\rm mol}({\rm ^{13}CO})$ is used as $\Sigma_{\rm mol}$, and thus the derived $\overline{W}_{\rm cloud}$ can be regarded as the average value on a 200 pc scale at each position.

Similarly, the vertical gravitational weight term associated with the atomic gas layer, $W_{\rm atom, 1kpc}$, is estimated on a 1~kpc scale as the sum of:

\begin{equation}
\begin{aligned}
W_{\mathrm{atom,1~kpc}} & =W_{\mathrm{atom,1~kpc}}^{\mathrm{self}}+W_{\mathrm{atom,1~kpc}}^{\mathrm{mol}}+W_{\mathrm{atom,1~kpc}}^{\mathrm{star}} \\
 & =\frac{\pi G}{2}\Sigma_{\mathrm{atom,1~kpc}}^{2}+\pi G\Sigma_{\mathrm{atom,1~kpc}}\Sigma_{\mathrm{mol,1~kpc}} \\
 & +\Sigma_{\mathrm{atom},1\mathrm{kpc}}\sqrt{2G\rho_{*,1\mathrm{kpc}}}\sigma_{\mathrm{atom}}.
\end{aligned}
\end{equation}
Here, $\sigma_{\text{atom}}$ and $\Sigma_{\rm atom, 1 kpc}$ are the vertical velocity dispersion of the atomic gas, for which we adopted a fixed value of 10~$\mathrm{km~s^{-1}}$ \citep{Leroy2008} and the atomic gas surface density, respectively.

Under vertical hydrostatic equilibrium, the corresponding equilibrium pressure is given by $P_{\rm DE} = \overline{W}_{\rm cloud} + W_{\rm atom,1kpc}$.

The $P_{\text{turb}}$ is proportional to the volume density of molecular gas ($\rho_{\text{mol}}$) and the square of its velocity dispersion ($\sigma_{\text{mol}}$). Since the mass surface density of molecular gas ($\Sigma_{\text{mol}}$) is directly obtained from the observations, we assumed the spherical molecular cloud with a uniform volume density and a diameter of molecular cloud $D_{\text{cloud}}$ \citep{Sun2020}. From the $\Sigma_{\text{mol}}$ and $\sigma_{\text{mol}}$ , the $P_{\text{turb}}$ is derived as:
\begin{equation}
P_{\mathrm{turb}}=\rho_{\mathrm{mol}}\sigma_{\mathrm{mol}}^2=\frac{3\Sigma_{\mathrm{mol}}\sigma_{\mathrm{mol}}^2}{2D_{\mathrm{cloud}}}.
\end{equation}
Here, we used  $\Sigma_{\text{mol}}(^{12}\text{CO})$ or $\Sigma_{\text{mol}}(^{13}\text{CO})$ as the $\Sigma_{\text{mol}}$.

\begin{table*}[h]
\centering
\caption{List of Additional Physical Properties}
\label{tab:additional_properties}
\begin{tabular}{p{4cm}p{3cm}p{3cm}p{3cm}}
\toprule
Definition & Symbol & Unit & Data source \\
\midrule
Atomic gas surface density & $\Sigma_{\mathrm{atom}}$ & $M_{\odot} \,\mathrm{pc}^{-2}$ & Koribalski et al. 2018 \\
Stellar mass surface density & $\Sigma_{*}$ & $M_{\odot} \,\mathrm{pc}^{-2}$ & Sheth et al. 2010 \\
Molecular gas surface density & $\Sigma_{\text{mol}}$ & $M_{\odot} \,\mathrm{pc}^{-2}$ & This work $^{12}$CO($J=2-1$)\\
Turbulent pressure in molecular gas & $P_{\mathrm{turb}}$ & $k_{\mathrm{B}} \,\mathrm{K} \,\mathrm{cm}^{-3}$ & This work \\
ISM equilibrium pressure & $P_{\mathrm{DE}}$ & $k_{\mathrm{B}} \,\mathrm{K} \,\mathrm{cm}^{-3}$ & All combined \\
\bottomrule
\end{tabular}
\end{table*}

$\Sigma_{\mathrm{atom,1kpc}}$ on a scale of 1 kpc was derived from the H~I 21~cm line data observed with ATCA \citep{Koribalski2018}. The 1~kpc spatial scale is adopted to represent the large-scale galactic disk environment relevant for evaluating the vertical dynamical equilibrium pressure of the ISM, since the dynamical equilibrium pressure reflects the vertical gravitational weight of the gas layer averaged over the galactic disk rather than the internal pressure of individual molecular clouds (e.g., \citealt{Sun2020}). These data are based on interferometric observations without additional single-dish zero-spacing information. Therefore, extended diffuse H~I emission on scales larger than the maximum recoverable scale may be partially resolved out, and the derived H~I surface densities should be regarded as lower limits. Since the synthesized beam size of the H~I data (1.8~kpc $\times$ 1.3~kpc) is larger than 1~kpc, we estimated a beam filling factor by assuming that the characteristic size of the H~I distribution within our regions is 1~kpc. We then derived the integrated intensity ($I_{\rm 21,1kpc}$) on a 1~kpc scale from the beam-diluted intensity in the original H~I map. We calculated $\Sigma_{\mathrm{atom,1kpc}}$ using a standard conversion factor that includes the mass contributions of helium and heavy elements \citep{Leroy2021}.

\begin{equation}
\Sigma_{\mathrm{atom,1kpc}}=1.97\times10^{-2}\times I_{21,1\mathrm{kpc}}\cos i\ (\rm M_{\odot} pc^{-2}).
\end{equation}
Here, the integration range for $I_{\mathrm{21,1kpc}}$ is from $221~\mathrm{kms^{-1}}$ to $427~\mathrm{kms^{-1}}$.

$\Sigma_{*, 1 \mathrm{kpc}}$ was calculated using Spitzer 3.6~$\mu$m data \citep{Sheth2010}. After smoothing the data to a 1~kpc scale by convolving a Gaussian kernel, $\Sigma_{*}$ was determined using the formula with a mass-to-light ratio $\Upsilon_{3.6} = 0.47$~$M_{\odot}$/$L_{\odot}$ \citep{McGaugh2014}:
\begin{equation}
\Sigma_{*, 1 \mathrm{kpc}}=3.3\times10^2\times I_{3.6,1\mathrm{kpc}}\cos i\ (\rm M_{\odot} pc^{-2}).
\end{equation}
Here, $I_{3.6,1\mathrm{kpc}}$ is the smoothed 3.6~$\mu$m intensity with units of $\mathrm{MJy~sr}^{-1}$.
The stellar volume density ($\rho_{*, 1 \mathrm{kpc}}$) is estimated from $\Sigma_{*, 1 \mathrm{kpc}}$ with the geometric structure of the stellar disk. Assuming an isothermal exponential vertical distribution and adopting a stellar disk flattening ratio $R_{*}/H_{*} = 7.3$ \citep{Kregel2002}, $\rho_{*, 1 \mathrm{kpc}}$ at the midplane can be calculated as
\begin{equation}
\rho_{*,1\mathrm{kpc}}=\frac{\Sigma_{*,1\mathrm{kpc}}}{4H_{*}}=\frac{\Sigma_{*,1\mathrm{kpc}}}{0.54R_{*}} (\rm M_{\odot} pc^{-3}).
\end{equation}
Here, $R_{*}$ is the scale length of stellar disk in M83. We adopted the value of $R_{*} = 2.9$~kpc from \citet{Leroy2008}.

Historically, it has been well established since the 1970s \citep[e.g.,][]{Larson1981, Solomon1987} that molecular clouds in the Milky Way are predominantly self-gravitating. Our motivation for evaluating the full $P_{\rm DE}$ equations is to quantify the relative importance of the molecular, stellar, and atomic gravitational terms across different environments in M83. To clarify which physical components dominate the total vertical weight term, we numerically evaluate each term in Equations (9) and (10). The fractional contributions are summarized in Table~\ref{tab:pressure_contributions}.

In the bar region, the molecular self-gravity term provides the dominant contribution (approximately 69\%), followed by the stellar gravity term (approximately 27\%). Similar to the bar region, in the spiral arm region, the molecular self-gravity term contribute most (64\%) and stellar gravity terms contribute 32\%. The contribution from the atomic gas layer is negligible ($\sim$1-2\%) in both environments. Even allowing for a factor due to possible missing large-scale H\,{\sc i} emission would not affect this result. This is consistent with the classical picture that cloud self-gravity remains the dominant term, while the stellar potential provides a non-negligible secondary contribution in the bar.

For the baseline estimation above, we adopted $D_{\rm cloud} = 70$~pc corresponding to the median value of molecular cloud sizes in M83 \citep{Hirota2024}. However, as pointed out by classic studies for the Milky Way, the mass-weighted average cloud diameter is typically smaller, around $\sim$40~pc \citep[e.g.,][]{Scoville1987}. The choice of $D_{\rm cloud}$ introduces a systematic uncertainty, but its effect can be evaluated analytically from Equations (9)--(11). In our formulation, $\overline{W}_{\rm cloud}^{\rm self}$ and $\overline{W}_{\rm cloud}^{\rm ext-mol}$ are independent of $D_{\rm cloud}$, whereas $\overline{W}_{\rm cloud}^{\rm star}$ scales linearly with $D_{\rm cloud}$ and $P_{\rm turb}$ scales as $D_{\rm cloud}^{-1}$.

Therefore, adopting a MW-like $D_{\rm cloud} = 40$~pc instead of 70~pc would reduce $\overline{W}_{\rm cloud}^{\rm star}$ by 42.9\% and increase $P_{\rm turb}$ by 75\%. Using the fractional contributions listed in Table~\ref{tab:pressure_contributions}, the corresponding decrease in the total $P_{\rm DE}$ is estimated to be 11.6\% in the bar and 13.7\% in the arm. Using the region-averaged fractional contributions, the corresponding ratio $P_{\rm turb}/P_{\rm DE}$ would increase by a factor of $\sim$1.98 in the bar and $\sim$2.03 in the arm. Thus, adopting a smaller fiducial cloud size would further strengthen the interpretation that the gas is more strongly turbulence-supported.

Because the molecular self-gravity term depends quadratically on the molecular gas surface density, variations in the adopted gas tracer (e.g., $^{12}$CO versus $^{13}$CO) directly propagate into the estimated equilibrium pressure. This naturally explains the difference between the pressure estimates shown in Figures~6 and 9.

\begin{table}
\centering
\caption{Numerical contributions to the dynamical equilibrium pressure.}
\label{tab:pressure_contributions}
\begin{tabular}{lccccc}
\toprule
Term & Fraction (Bar) & Fraction (Arm) \\
\midrule
$\overline{W}_{\rm cloud}^{\rm self}$ & 69.33\% & 63.98\% \\
$\overline{W}_{\rm cloud}^{\rm ext-mol}$  & 2.36\% & 2.08\% \\
$\overline{W}_{\rm cloud}^{\rm star}$            & 27.09\% & 31.93\% \\
$W_{\mathrm{atom,1~kpc}}^{\mathrm{self}}$            & 0.11\% & 0.06\% \\
$W_{\mathrm{atom,1~kpc}}^{\mathrm{mol}}$          & 0.17\% & 0.11\% \\
$W_{\mathrm{atom,1~kpc}}^{\mathrm{star}}$           & 1.03\% & 1.84\% \\
\midrule
Total $P_{\rm DE}$ & 100\% & 100\% \\
\bottomrule
\end{tabular}
\end{table}

\subsubsection{Pressure Ratio and SFE}
\begin{figure*}
 \begin{center}
  \includegraphics[width=\linewidth]{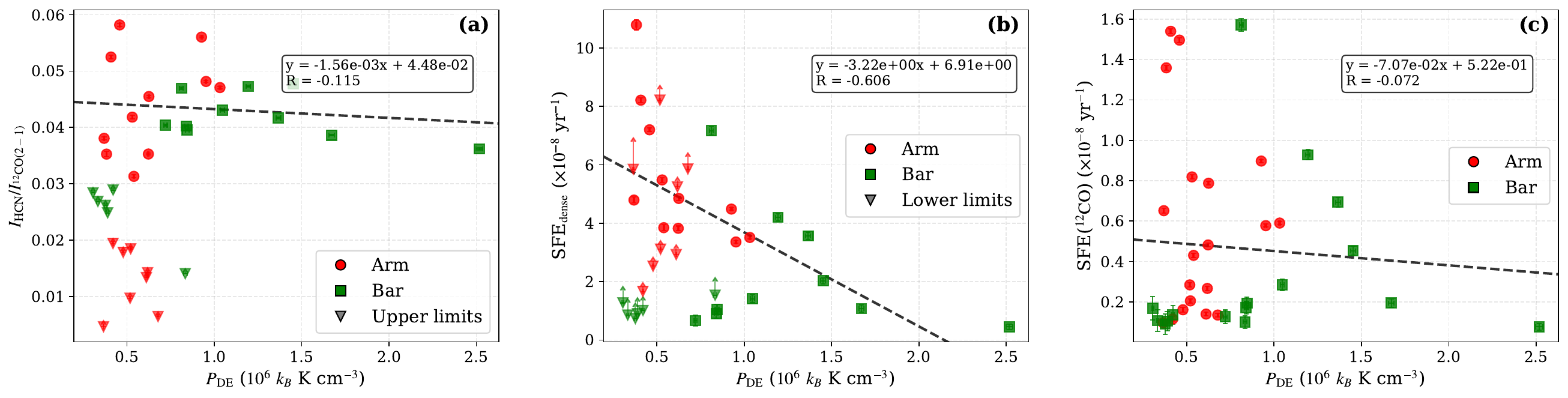}
 \end{center}
\caption{Relationships between (a) $ I_{\mathrm{HCN}} / I_{^{12}\mathrm{CO}(2-1)} $ ratio,  (b) SFE$_{\mathrm{dense}}$, and (c) SFE($^{12}\mathrm{CO}$) as a function of $P_{\mathrm{DE}}$ in M83. Data points for the bar and spiral arm regions are shown as green squares and red circles, respectively. The inverted triangles indicate data points with upper limits on the $ I_{\mathrm{HCN}} / I_{^{12}\mathrm{CO}(2-1)} $ in (a) and lower limits on the SFE$_{\mathrm{dense}}$ in (b). R indicates the value of the Pearson correlation coefficient. The SFE$_{\mathrm{dense}}$ in the bar region is systematically lower than in the arm region across nearly the entire pressure range, indicating suppressed efficiency of star formation from dense gas within the bar. \\
Alt text: Three scatter plots arranged side by side, each plotting a different parameter against the $P_{\text{DE}}$. green squares represent the bar region, and red circles represent the spiral arm region.}
\label{fig:Pde_parameters}
\end{figure*}

Figure~\ref{fig:Pde_parameters} shows that the correlations with $P_{\rm DE}$ alone are weak. We therefore do not consider $P_{\rm DE}$ itself to be the primary driver of the suppressed SFE, but use it mainly as a reference level when evaluating $P_{\rm turb}/P_{\rm DE}$.

\begin{figure*}
 \begin{center}
  \includegraphics[width=\linewidth]{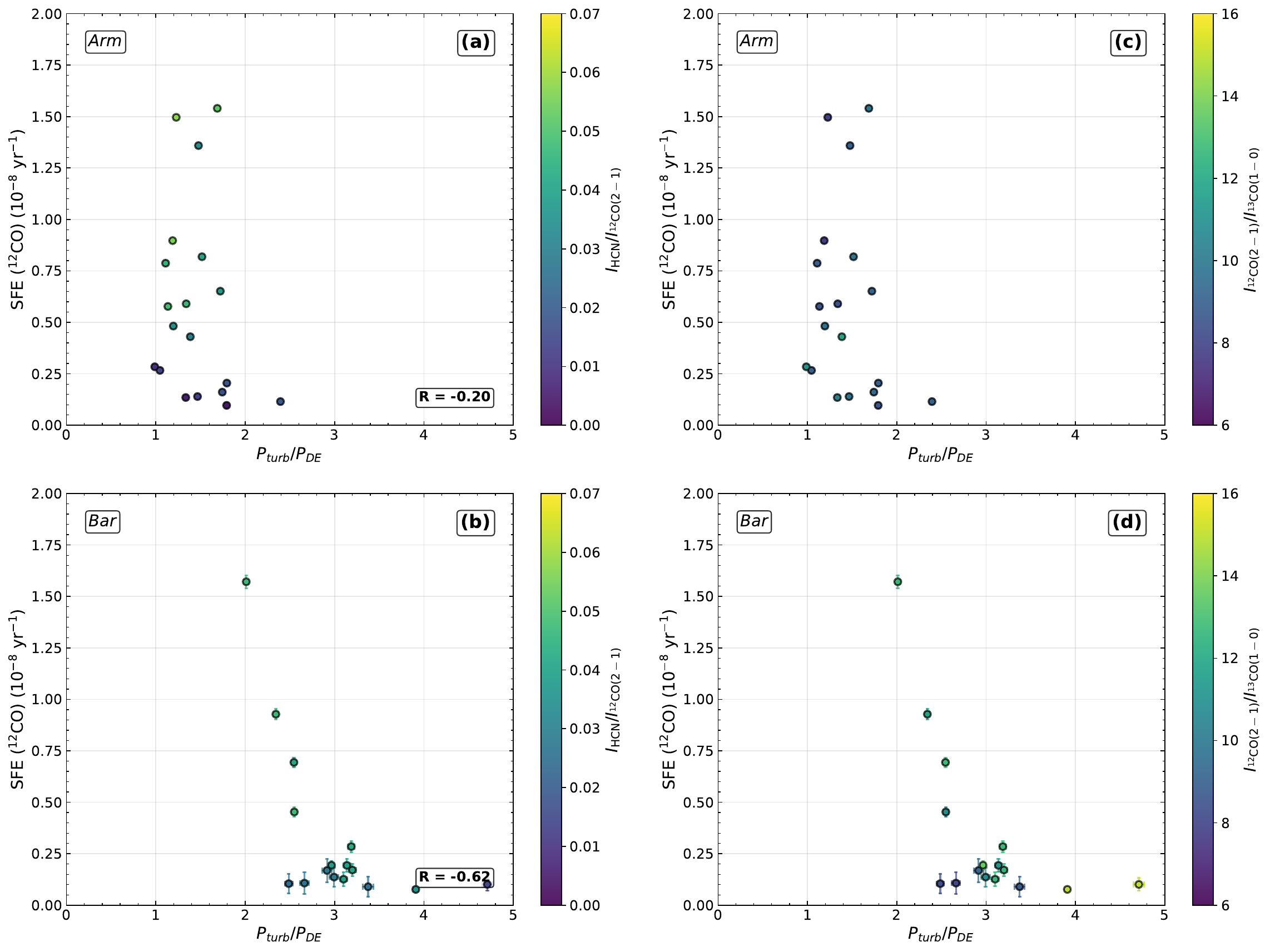}
 \end{center}
\caption{Relationship between turbulent-to-equilibrium pressure ratio ($P_{\mathrm{turb}} / P_{\mathrm{DE}}$) and SFE in M83. The upper two panels show the plots for the arm, while the lower two panels show those for the bar. The colors in (a) and (b) represent the $ I_{\mathrm{HCN}} / I_{^{12}\mathrm{CO}(2-1)} $ ratio, while those in panels (c) and (d) represent the $ I_{^{12}\mathrm{CO}(2-1)} / I_{^{13}\mathrm{CO}(1-0)} $ ratio. R indicates the value of the Pearson correlation coefficient. A negative correlation is observed in the bar region, indicating that higher turbulent pressure relative to equilibrium pressure corresponds to lower SFE. \\
Alt text: Four scatter plots arranged vertically, showing turbulent-to-equilibrium pressure ratio versus star formation efficiency. The top panel displays data for the arm region, and the bottom panel for the bar region.}
\label{fig:pressure_ratio}
\end{figure*}

Figure~\ref{fig:pressure_ratio} shows the relationship between the ratio of molecular gas turbulent pressure to dynamical equilibrium pressure ($P_{\mathrm{turb}} / P_{\mathrm{DE}}$) and the SFE($^{12}$CO). The data points are color–coded by the integrated intensity ratio of $ I_{\mathrm{HCN}} / I_{^{12}\mathrm{CO}(2-1)} $, and the ratio of $I_{^{12}\mathrm{CO}(2-1)} / I_{^{13}\mathrm{CO}(1-0)}$. In the arm regions, the $P_{\mathrm{turb}}/P_{\mathrm{DE}}$ value lies in the range of 1--2.4, while in the bar region it spans a wider range of 2--4.8. SFE($^{12}$CO) exhibits large scatter in the arm regions. In contrast, in the bar region, SFE($^{12}$CO) is mostly concentrated in the range of $(0-0.25) \times 10^{-8}\mathrm{yr^{-1}}$ except for a few position. 

As indicated by the large observed $\sigma_{\text{v}}$ values, it is clear that $P_{\mathrm{turb}}$ is the dominant pressure component in the system, indicating that turbulent support dominates the local dynamical state of the molecular gas. In the bar region, the systematically higher $P_{\mathrm{turb}}/P_{\mathrm{DE}}$ ratios imply stronger turbulent motions, which can hinder the formation of gravitationally bound structures and thereby contribute to the suppression of SFE.

Figure \ref{fig:pressure_ratio} (a) and (b) show the distribution of the $ I_{\mathrm{HCN}} / I_{^{12}\mathrm{CO}(2-1)} $ ratio which is an indicator of the dense gas fraction. In the arm, a tendency for the $ I_{\mathrm{HCN}} / I_{^{12}\mathrm{CO}(2-1)} $ ratio to increase with rising SFE is seen, supporting the idea that dense gas contributes efficiently to star formation. In contrast, in the bar structure, a high $ I_{\mathrm{HCN}} / I_{^{12}\mathrm{CO}(2-1)} $ ratio is maintained even in regions with low SFE, indicating a peculiar state where dense gas is abundant but its conversion to star formation is suppressed. This result strongly suggests that turbulent pressure is a primary mechanism hindering star formation in the bar structure.

In Section 4.1, we have pointed out the trend that star formation is suppressed in molecular gas with a high dense-gas fraction in the bar. \citet{Neumann2024} also reported a discrepancy in the bar region where the $ I_{\mathrm{HCN}} / I_{^{12}\mathrm{CO}(2-1)} $ ratio is similar to the disk but the SFR/HCN ratio is low, suggesting this phenomenon is not unique to a single galaxy.

In Figure \ref{fig:pressure_ratio} (c) and (d) , the colors of the plots indicate the values of the $I_{^{12}\mathrm{CO}(2-1)} / I_{^{13}\mathrm{CO}(1-0)}$ ratio. The bar region shows systematically higher ratios than those in the arm region. In addition, we found that the ratio tends to be higher in regions with low SFE($^{12}\mathrm{CO}$).

This spatial variation suggests differences in the physical condition or chemical composition of molecular gas between the two regions. The elevated $I_{^{12}\mathrm{CO}(2-1)} / I_{^{13}\mathrm{CO}(1-0)}$ ratio observed in the bar region could be attributed to two primary factors: a lower abundance of $^{13}$CO or a lower optical depth of the $^{13}$CO($J=1-0$) line. The mechanism of selective photodissociation, which can preferentially destroy $^{13}\mathrm{CO}$, was originally proposed to explain isotopologue ratios in interstellar clouds \citep{van1988}. In the context of nearby galaxies, such selective photodissociation has been discussed as a potential driver of variations in $^{12}\mathrm{CO}$/$^{13}\mathrm{CO}$ ratios \citep[e.g.,][]{Davis2014}. However, this mechanism appears unlikely to explain the lower ratio in our study of M83 because the observed regions of the bar and spiral arms are sufficiently distant from the nuclear region of M83. Furthermore, while photodissociation by UV photons from star formation could be another mechanism, the lower star formation activity in the bar region also rules out this possibility.

\begin{figure}
 \begin{center}
  \includegraphics[width=0.9\linewidth]{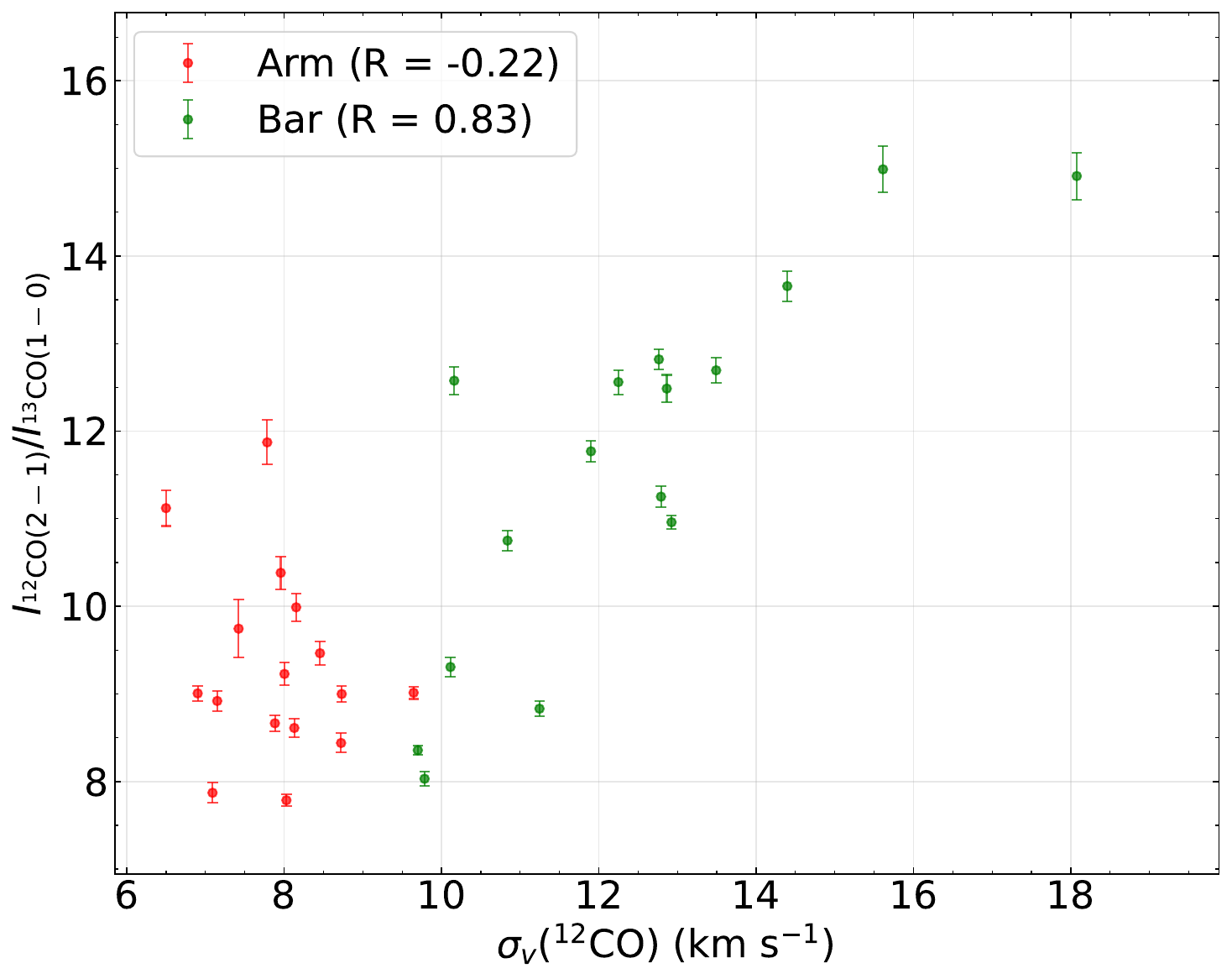}
 \end{center}
\caption{The relationship between the molecular line with of $^{12}$CO($J=2-1$) ($\sigma_v\mathrm{(^{12}CO)}$) and $I_{^{12}\mathrm{CO}(2-1)} / I_{^{13}\mathrm{CO}(1-0)}$ ratios in the arm(red) and bar(green) regions. R indicates the value of the Pearson correlation coefficient.\\
Alt text: A scatter plot showing data points for the spiral arm and bar regions, plotting molecular line width against the ratio of two CO isotopologue intensities.}
\label{fig:sigma}
\end{figure}

On the other hand, the $^{13}\mathrm{CO}$($J=1-0$) line may be optically thin in the bar. Figure~\ref{fig:sigma} shows a trend where the line width increases with the $^{12}\mathrm{CO}$($J=2-1$)/$^{13}\mathrm{CO}$($J=1-0$) ratio. On molecular cloud scales, it is known that a larger internal velocity dispersion reduces line overlapping, leading to a lower optical depth. The broad velocity dispersion observed in the bar structure supports the interpretation that the $^{13}$CO($J=1-0$) line is in a more optically thin condition than in the arm.

\citet{Sorai2012} revealed that the molecular gas in the bar region is in a gravitationally unbound state, exhibiting a large inter-cloud velocity dispersion and strong non-circular motions ($\sim$100 km s$^{-1}$). In such an environment, the integrated intensity increases because the $^{12}$CO($J=2-1$) emission remains optically thick,  and the linewidth of spectrum broadens with the velocity dispersion. In contrast, since $^{13}$CO($J=1-0$) is intrinsically optically thin, its integrated intensity does not increase in the same manner. Moreover, the integrated intensity of $^{13}$CO($J=1-0$) can be underestimated due to low S/N as the increased velocity dispersion can further reduce the optical depth and weakens the line emission. As a consequence, the $I_{^{12}\mathrm{CO}(2-1)} / I_{^{13}\mathrm{CO}(1-0)}$ ratio becomes larger in regions with broader line widths.

Therefore, the high $\sigma_{^{12}\mathrm{CO}}$ can be interpreted as a consistent observational manifestation of a physical environment where vigorous inter-cloud motions dominate and the gas is gravitationally unbound state. Under such an environment, the $P_{\mathrm{turb}}/P_{\mathrm{DE}}$ ratio of the molecular gas increases, leading to low SFE.

\subsection{HCN Luminosity and IR Luminosity}
We compare the relationship between the HCN luminosity ($L_{\text{HCN}}$) and the IR luminosity ($L_{\mathrm{IR}}$) in M83 with the universal scaling relation established in nearby galaxies. Since $L_{\mathrm{IR}}$ measurements on a scale of 200 pc are not available for M83, we estimate the $L_{\mathrm{IR}}$ from the SFR derived from the extinction-corrected H$\alpha$ intensity. We converted $I_{\text{HCN}}$ into $L_{\text{HCN}}$ using the formula following \citet{Leroy2022}:
\begin{equation}
    L_{\rm HCN} = I_{\mathrm{HCN}} \times A~(\mathrm{K~km~s^{-1}~pc^{2}})
\end{equation}
where $A$ is the beam area in the unit of pc$^2$.

$L_{\mathrm{IR}}$ is estimated from the SFR obtained from H$\alpha$ through the formula \citep{Kennicutt2012}:
\begin{equation}
    \mathrm{SFR}~(\mathrm{M_\odot\,yr^{-1}}) = 1.27 \times 10^{-43} \times L_{\mathrm{IR}}~(L_{\odot}).
\end{equation}
Here, we assumed that the SFR estimated from H$\alpha$ is the same as that derived from the IR luminosity. 

\begin{figure}
 \begin{center}
  \includegraphics[width=\linewidth]{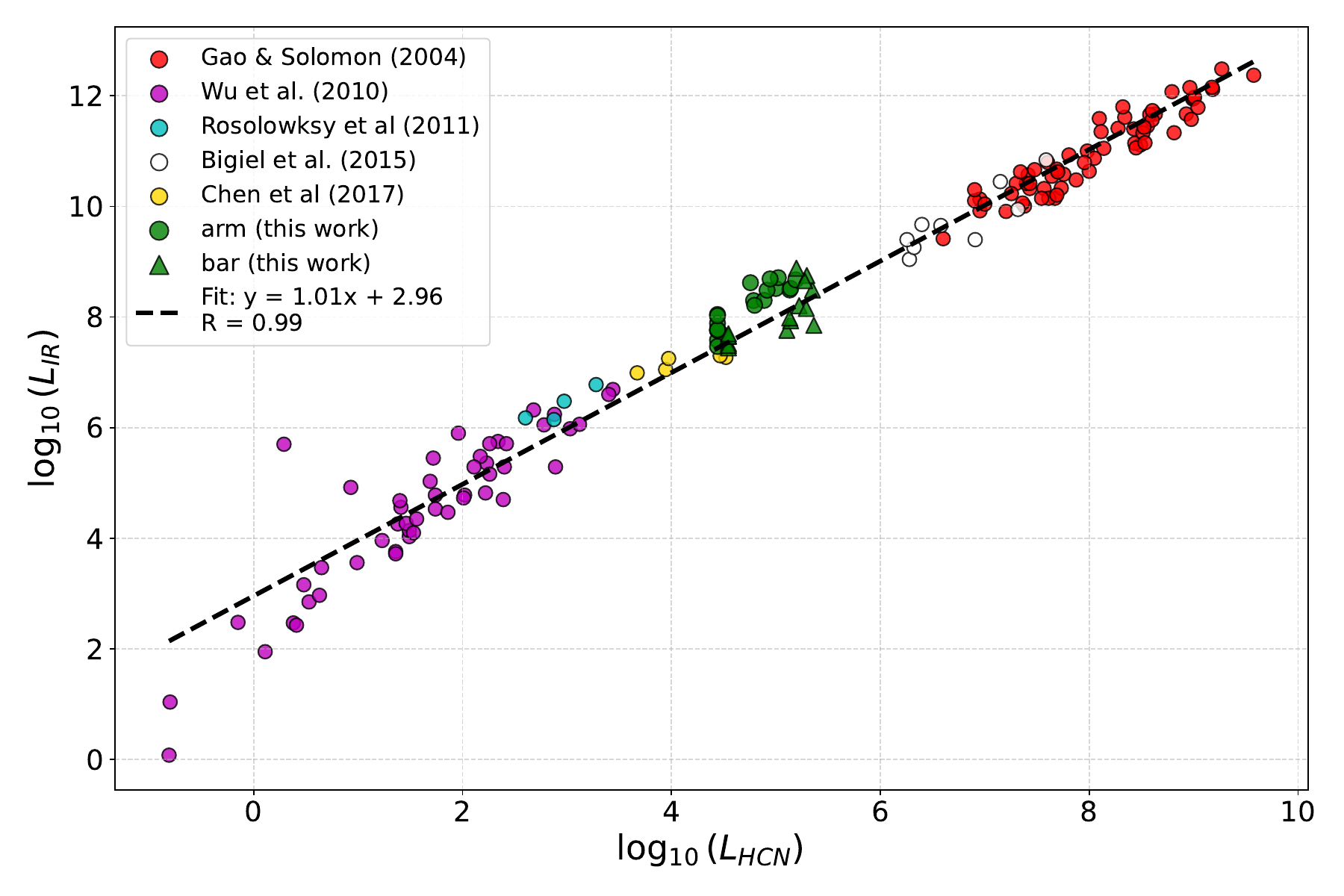}
 \end{center}
\caption{Relationship between the HCN luminosity ($L_{\rm HCN}$) and the infrared luminosity $L_{\text{IR}}$. We estimated $L_{\text{IR}}$ for M83 from the SFR derived from the extinction-corrected H$\alpha$ intensity using Equation (16). The dashed line shows a scaling relation fitted to the all data in this Figure, following the form of the universal relation proposed by \citet{Gao2004}. R indicates the value of the Pearson correlation coefficient. Data points in the arm and the bar of M83 are shown as green circles and green triangles, respectively. \\
Alt text: A scatter plot showing the relationship between two luminosity measures. Green circular data points represent spiral arm regions, green triangular points represent bar regions, and a dashed line indicates a fitted scaling relation.}
\label{fig:dense_gas_sfr}
\end{figure}

Figure~\ref{fig:dense_gas_sfr} shows the relationship between $L_{\mathrm{HCN}}$ and $L_{\mathrm{IR}}$ estimated from our analysis as well as the data from other works \citep[][]{Gao2004, Wu2010, Rosolowsky2011, Bigiel2015, Chen2017}. In Figure 8, both the bar and arm data are distributed along a dashed line, as proposed by \citet{Gao2004}. The bar data points tend to lie slightly below this line, reflecting the lower $\mathrm{SFE_{dense}}$ in bars compared to arms. Significant scatter in the $L_{\mathrm{HCN}}$–$L_{\mathrm{IR}}$ relation has been recognized over a wide range of spatial scales \citep[e.g.,][]{Jiménez2019}. The scatter is thought to arise from variations in the $\mathrm{SFE_{dense}}$ under different environmental factors such as stellar mass surface density, molecular gas mass surface density, and $\mathrm{P_{DE}}$. In addition, our results suggest that the influence of gas dynamics in bars can also contribute to the scatter in the $L_{\mathrm{HCN}}$–$L_{\mathrm{IR}}$ relation.

\section{Summary}

We investigated the SFE in the bar region of the nearby barred spiral galaxy M83 and its physical causes, using multi-wavelength observations on a spatial scale of $\sim$200 pc. The main results and conclusions are as follows.

\begin{enumerate}
\item We confirmed that the SFE is suppressed in the bar region compared with the arm region. Although the molecular gas surface densities traced by either $^{12}$CO($J=2-1$) or $^{13}$CO($J=1-0$) is comparable to or higher than those in the spiral arm region, the star formation rate surface density shows no corresponding enhancement in the bar. This indicates that the SFE of the moderate density gas is systematically lower in the bar.

\item Furthermore, the dense gas SFE$_{\rm dense}$, traced by HCN($J = 1-0$), is also suppressed in the bar. For a given dense gas surface density, the star formation rate is lower in the bar than in the arm. This implies that the SFE of dense gas itself is suppressed in the bar, suggesting that the suppression mechanism the SFE also affects the final stages of the star formation processes from the dense molecular gas to the protostars.

\item The suppression of both SFE and SFE$_{\rm dense}$ can be understood in terms of the dynamically harsh environment in the bar. The systematically larger CO line widths indicate stronger shocks, shear, and cloud--cloud collisions, which hinder molecular cloud contraction and core growth. The higher $P_{\rm turb}/P_{\rm DE}$ ratios in the bar are consistent with this picture, providing a quantitative measure of the enhanced turbulent state.
\end{enumerate}

In summary, our multi-molecular line observations of M83 reveal a physical scenario in which the unique dynamical environment of the bar reduces the efficiency of the entire process that converts molecular gas into stars. The galactic structures, such as bars, can significantly regulate SFE through the dynamical effects, and their effects extend beyond the accumulation of molecular gas to the processes from the dense gas to the protostellar formation.

\begin{ack}
The authors thank the anonymous referee for useful suggestions that improved the manuscript, and the editor Kazuyuki Muraoka. This paper makes use of the ALMA dataset ADS/JAO.ALMA\#2013.1.00889.S. ALMA is a partnership of the ESO (representing its member states), the NSF (USA) and NINS (Japan), together with the NRC (Canada) and the NSC and ASIAA (Taiwan), in cooperation with the Republic of Chile. The Joint ALMA Observatory is operated by the ESO, the AUI/NRAO and the NAOJ. The authors are grateful to the ALMA staff for their excellent support.  This study is supported by a Grant-in-Aid from the Ministry of Education, Culture, Sports, Science, and Technology of Japan (No.25K01042).  Y.W. also acknowledges support from a Grant-in-Aid from the Ministry of Education, Culture, Sports, Science, and Technology of Japan (No.24K00675, 25H00676).
\end{ack}
 

\appendix 
\section{Analysis of Turbulent Pressure and Equilibrium Pressure Based on $^{13}$CO($J=1-0$)}
\label{app:a}

We calculated the turbulent pressure ($P_{\rm turb}(^{13}{\rm CO})$) and the dynamical pressure ($P_{\rm DE}(^{13}{\rm CO})$) using $^{13}$CO($J=1-0$) data instead of $^{12}$CO($J=2-1$) data. The analysis method is identical to that in Section~4.3 of the main text.

\begin{figure*}
 \begin{center}
  \includegraphics[width=0.6\linewidth]{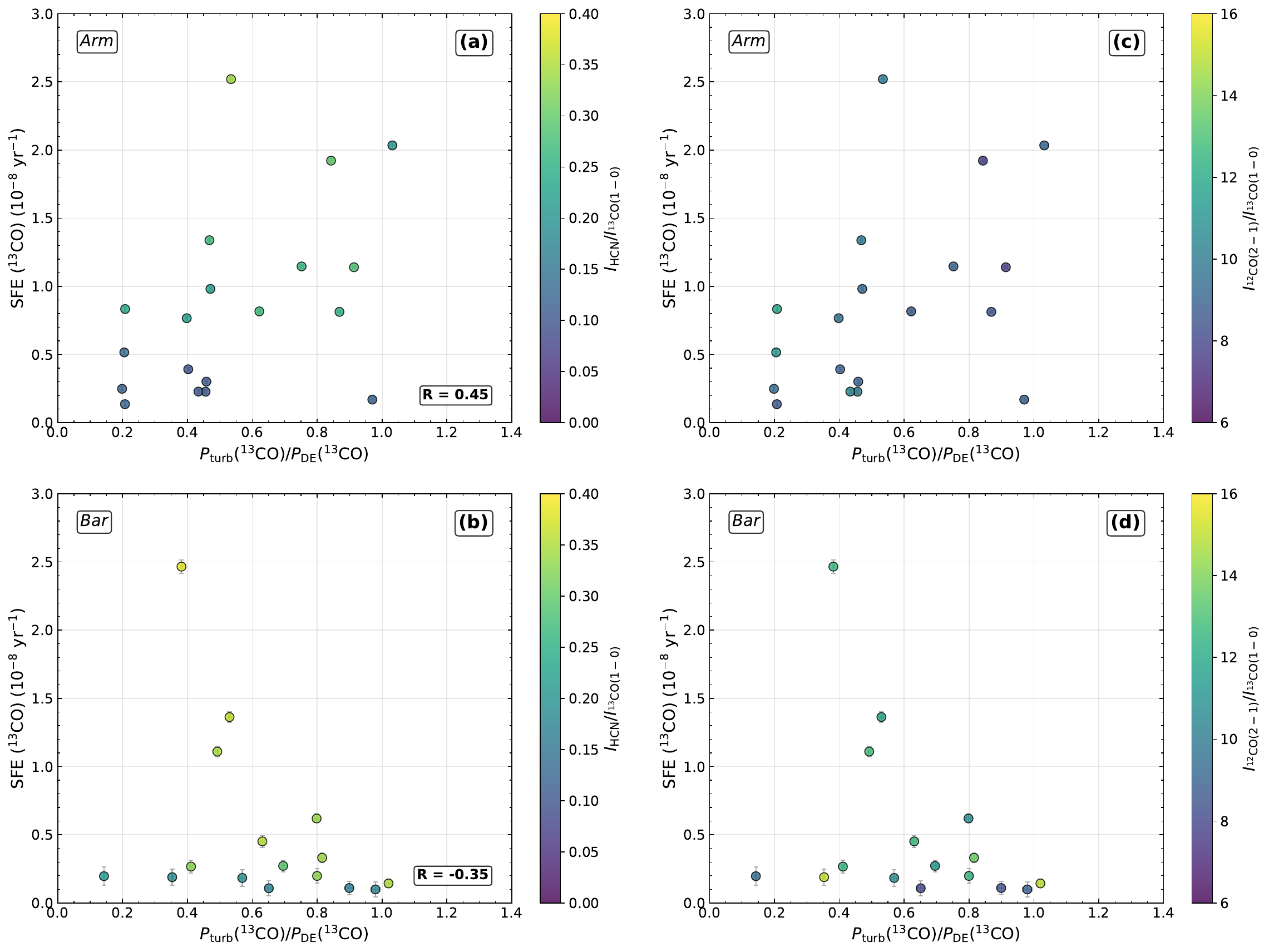}
 \end{center}
\caption{Relationship between turbulent-to-equilibrium pressure ratio ($P_{\mathrm{turb}} / P_{\mathrm{DE}}$) derived from $^{13}$CO($J=1-0$) data and SFE in M83. The upper two panels show the plots for the arm, while the lower two panels show those for the bar. The colors in (a) and (b) represent the $ I_{\mathrm{HCN}} / I_{^{12}\mathrm{CO}(2-1)} $ ratio, while those in panels (c) and (d) represent the $I_{^{12}\mathrm{CO}(2-1)} / I_{^{13}\mathrm{CO}(1-0)}$ ratio. R indicates the value of the Pearson correlation coefficient.\\
Alt text: Four scatter plots arranged vertically, showing turbulent-to-equilibrium pressure ratio versus star formation efficiency. The top panel displays data for the arm region, and the bottom panel for the bar region.}
\label{fig:13C_pressure_ratio}
\end{figure*}

\begin{figure*}
 \begin{center}
  \includegraphics[width=0.5\linewidth]{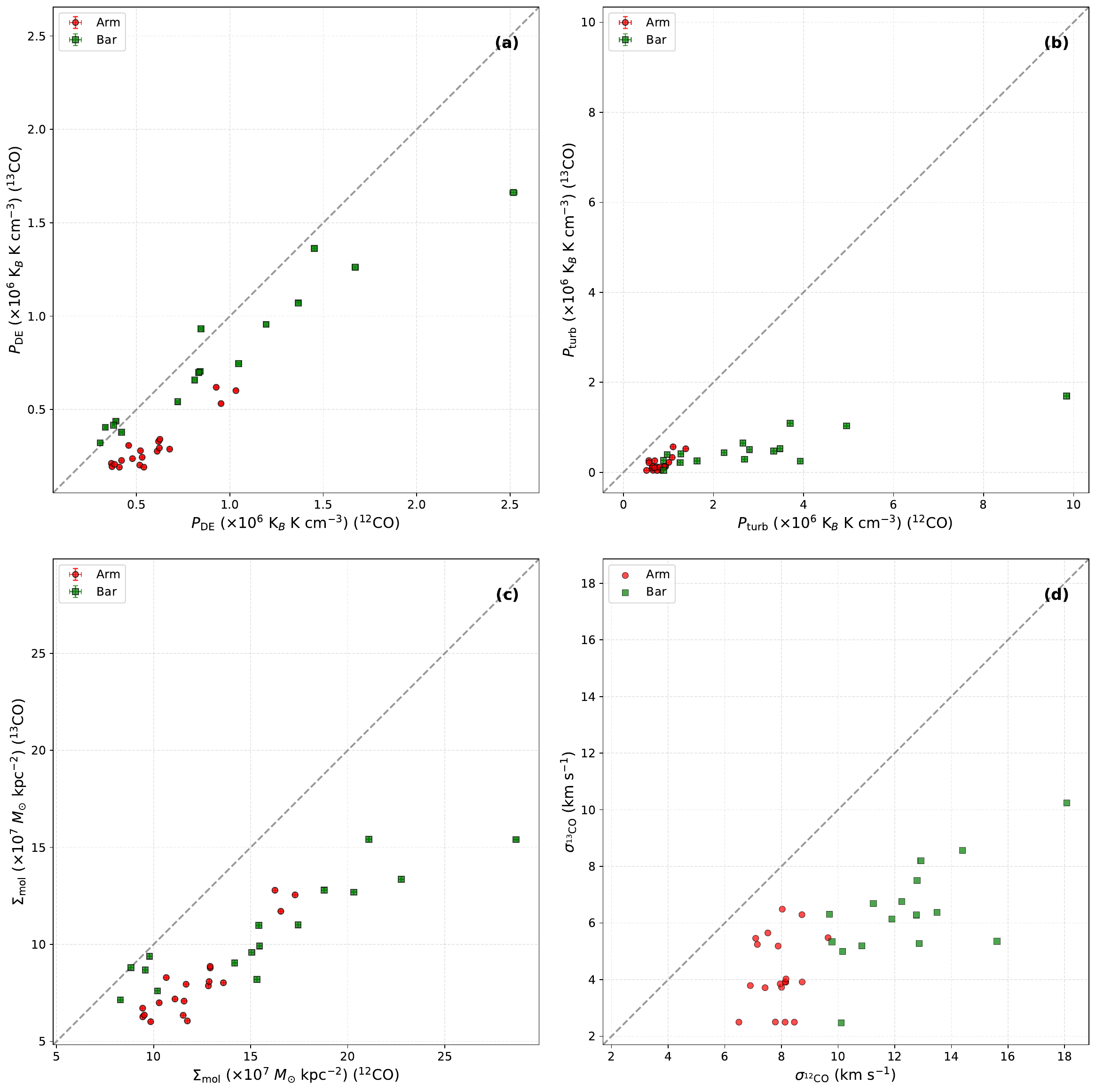}
 \end{center}
\caption{Comparison of physical parameters derived from $^{12}$CO($J=2-1$)\ and $^{13}$CO($J=1-0$): (a) dynamical equilibrium pressure $P_{\mathrm{DE}}$, (b) turbulent pressure $P_{\mathrm{turb}}$, (c) molecular gas surface density $\Sigma_{\mathrm{mol}}$, and (d) velocity dispersion $\sigma_{\mathrm{mol}}$. Bar and arm regions are represented in red and blue, respectively. This comparison elucidates the systematic differences in the $P_{\mathrm{turb}}/P_{\mathrm{DE}}$ ratio between $^{12}$CO($J=2-1$)\ and $^{13}$CO($J=1-0$)\ discussed in Appendix~\ref{app:a}. \\
Alt text: A four-panel comparison of derived physical parameters from two molecular line tracers. Red indicates the bar region, and blue indicates the spiral arm region in each panel.}
\label{fig:parameter_comparison}
\end{figure*}

Figure \ref{fig:13C_pressure_ratio} shows correlation plot of the $P_{\rm turb}(^{13}{\rm CO}) / P_{\rm DE}(^{13}{\rm CO})$ ratios and $\mathrm{SFE}(^{13}{\rm CO})$, which is equivalent to Figure \ref{fig:pressure_ratio} in Section 4.3.  We confirmed that the trends are in very good agreement those derived from $^{12}$CO($J=2-1$) data (Figure \ref{fig:pressure_ratio}). The main conclusion that a negative correlation exists between $P_{\mathrm{turb}} / P_{\mathrm{DE}}$ and SFE in the bar region is also supported by using $^{13}$CO($J=1-0$) data.

As shown in Figure \ref{fig:13C_pressure_ratio}, the primary trend of a negative correlation between $P_{\mathrm{turb}}/P_{\mathrm{DE}}$ and $\mathrm{SFE}(^{13}{\rm CO})$ in the bar region is in good agreement with the results obtained using $^{12}$CO($J=2-1$)\ (Figure \ref{fig:pressure_ratio} in the main text).

However, the absolute values of $P_{\mathrm{turb}}/P_{\mathrm{DE}}$ calculated from $^{13}$CO($J=1-0$)\ were systematically lower than those estimated from $^{12}$CO($J=2-1$) by approximately $70\%$ (Figure 10). This systematic difference stems from the differences in the molecular gas surface density and the velocity dispersion derived from the two tracers. The $\Sigma_{\mathrm{mol}}$ derived from $^{13}$CO($J=1-0$)\ is systematically lower than that derived from $^{12}$CO($J=2-1$). Since the calculation of $P_{\mathrm{DE}}$ strongly depends on $\Sigma_{\mathrm{mol}}$ according to formula ~(9) and (10), this difference is the main factor leading to a lower $P_{\mathrm{DE}}$ based on $^{13}$CO($J=1-0$)\ (Figure \ref{fig:parameter_comparison}a). In addition, the velocity dispersion $\sigma_{\mathrm{mol}}$ measured from $^{13}$CO($J=1-0$)\ is systematically smaller than that measured from $^{12}$CO($J=2-1$). As shown in formula~(11) ($P_{\mathrm{turb}} \propto \Sigma_{\mathrm{mol}} \sigma_{\mathrm{mol}}^{2}$), $P_{\mathrm{turb}}$ is proportional to the square of $\sigma_{\mathrm{mol}}$. Therefore, this difference in $\sigma_{\mathrm{mol}}$ is the decisive factor causing a smaller $P_{\mathrm{turb}}$ based on $^{13}$CO($J=1-0$)\ (Figure \ref{fig:parameter_comparison}b).

To further clarify the physical origin of the difference between the pressure estimates derived from $^{12}$CO and $^{13}$CO, we also numerically evaluated the fractional contribution of each term in Equations (9) and (10) using the $^{13}$CO-based $\Sigma_{\rm mol}$. The results are summarized in Table~\ref{tab:pressure_contributions_13co}. In both the arm and bar regions, the dominant contributions to $P_{\rm DE}$ arise from the molecular self-gravity term ($\overline{W}_{\rm cloud}^{\rm self}$) and the stellar gravity term ($\overline{W}_{\rm cloud}^{\rm star}$). The contribution from the extended molecular disk term remains small ($\sim2\%$), and the atomic-gas-related terms collectively contribute less than $\sim2\%$ of the total pressure. This confirms that the atomic gas layer plays only a minor role in determining the dynamical equilibrium pressure in our target regions. A notable difference between the arm and bar environments is the relative importance of the stellar gravity term. In the arm region, the molecular self-gravity term contributes about $55\%$ of the total pressure, while the stellar gravity term contributes about $41\%$. In contrast, in the bar region the stellar gravity term becomes the dominant contributor ($\sim58\%$), exceeding the molecular self-gravity contribution ($\sim39\%$). This behavior can be understood as a consequence of the lower molecular gas surface densities derived from $^{13}$CO compared to $^{12}$CO. Since the molecular self-gravity term scales with $\Sigma_{\rm mol}^{2}$, a reduction in $\Sigma_{\rm mol}$ significantly decreases this term, thereby increasing the relative importance of the stellar gravitational potential. This effect explains why the $P_{\rm DE}$ values derived from $^{13}$CO are systematically lower than those derived from $^{12}$CO, as shown in Figure~\ref{fig:parameter_comparison}(a).

\begin{table}
\centering
\caption{Fractional contributions to the dynamical equilibrium pressure derived from $^{13}$CO($J=1-0$).}
\label{tab:pressure_contributions_13co}
\begin{tabular}{lcc}
\toprule
Term & Fraction (Bar) & Fraction (Arm) \\
\midrule
$\overline{W}_{\rm cloud}^{\rm self}$ & 38.72\% & 55.07\% \\
$\overline{W}_{\rm cloud}^{\rm ext-mol}$ & 1.91\% & 2.72\% \\
$\overline{W}_{\rm cloud}^{\rm star}$ & 58.43\% & 40.51\% \\
$W_{\rm atom}^{\rm self}$ & 0.02\% & 0.12\% \\
$W_{\rm atom}^{\rm mol}$ & 0.08\% & 0.22\% \\
$W_{\rm atom}^{\rm star}$ & 0.84\% & 1.36\% \\
\midrule
Total $P_{\rm DE}$ & 100\% & 100\% \\
\bottomrule
\end{tabular}
\end{table}

As discussed in Section 4.3.1, adopting a MW-like $D_{\rm cloud} = 40$~pc instead of 70~pc decreases $\overline{W}_{\rm cloud}^{\rm star}$ by 42.9\% and increases $P_{\mathrm{turb}}$ by 75\%. Under this assumption, the molecular self-gravity and stellar gravity contributions become comparable even in the bar region for the $^{13}$CO-based estimates, leading to an even higher $P_{\mathrm{turb}}/P_{\mathrm{DE}}$ ratio. This further demonstrates that the turbulence-supported nature of the bar gas is a robust conclusion, independent of the chosen molecular gas tracer and assumed cloud size.

Nevertheless, despite the differences in the derived absolute values, the relative trends between different environments (i.e., higher $P_{\mathrm{turb}}/P_{\mathrm{DE}}$ and lower SFE in the bar region) are consistent for both $^{12}$CO($J=2-1$) and $^{13}$CO($J=1-0$). This strongly supports the robustness of our conclusions of this study.

\end{document}